\input harvmac.tex



\let\includefigures=\iftrue
\newfam\black
\includefigures
\input epsf
\def\figin{\epsfcheck\figin}\def\figins{\epsfcheck\figins}
\def\epsfcheck{\ifx\epsfbox\UnDeFiNeD
\message{(NO epsf.tex, FIGURES WILL BE IGNORED)}
\gdef\figin##1{\vskip2in}\gdef\figins##1{\hskip.5in}
\else\message{(FIGURES WILL BE INCLUDED)}%
\gdef\figin##1{##1}\gdef\figins##1{##1}\fi}
\def\DefWarn#1{}
\def\figinsert{\goodbreak\midinsert}
\def\ifig#1#2#3{\DefWarn#1\xdef#1{fig.~\the\figno}
\writedef{#1\leftbracket fig.\noexpand~\the\figno}%
\figinsert\figin{\centerline{#3}}\medskip\centerline{\vbox{\baselineskip12pt \advance\hsize by
-1truein\noindent\footnotefont{\bf Fig.~\the\figno:} #2}}
\bigskip\endinsert\global\advance\figno by1}
\else
\def\ifig#1#2#3{\xdef#1{fig.~\the\figno}
\writedef{#1\leftbracket fig.\noexpand~\the\figno}%
\global\advance\figno by1} \fi

\font\cmss=cmss10 \font\cmsss=cmss10 at 7pt

\def\IB{\relax\hbox{$\inbar\kern-.3em{\rm B}$}}
\def\IC{\relax\hbox{$\inbar\kern-.3em{\rm C}$}}
\def\IQ{\relax\hbox{$\inbar\kern-.3em{\rm Q}$}}
\def\ID{\relax\hbox{$\inbar\kern-.3em{\rm D}$}}
\def\IE{\relax\hbox{$\inbar\kern-.3em{\rm E}$}}
\def\IF{\relax\hbox{$\inbar\kern-.3em{\rm F}$}}
\def\IG{\relax\hbox{$\inbar\kern-.3em{\rm G}$}}
\def\IGa{\relax\hbox{${\rm I}\kern-.18em\Gamma$}}
\def\IH{\relax{\rm I\kern-.18em H}}
\def\IK{\relax{\rm I\kern-.18em K}}
\def\IL{\relax{\rm I\kern-.18em L}}
\def\IP{\relax{\rm I\kern-.18em P}}
\def\IR{\relax{\rm I\kern-.18em R}}
\def\Z{\relax\ifmmode\mathchoice
{\hbox{\cmss Z\kern-.4em Z}}{\hbox{\cmss Z\kern-.4em Z}} {\lower.9pt\hbox{\cmsss Z\kern-.4em Z}}
{\lower1.2pt\hbox{\cmsss Z\kern-.4em Z}}\else{\cmss Z\kern-.4em Z}\fi}
\def\IZ{Z\!\!\!Z}
\def\II{\relax{\rm I\kern-.18em I}}
\def\one{\relax{\rm 1\kern-.25em I}}

\def\S{{\bf S}}

\def\AdS{{\rm AdS}}
\def\Sp{{\rm Sp}}

\def\CLL{\relax{\CL\kern-.74em \CL}}

\def\CA {{\cal A}}

\def\CD {{\cal D}}

\def\CF {{\cal F}}
\def\CG {{\cal G}}
\def\CH {{\cal H}}

\def\CJ {{\cal J}}
\def\CK {{\cal K}}
\def\CL {{\cal L}}
\def\CM {{\cal M}}

\def\CS {{\cal S}}
\def\CT {{\cal T}}


\def\p{\partial}

\def\tilde{\widetilde}
\def\hat{\widehat}
\def\bar{\overline}

\def\dim{{\mathop{\rm dim}}}

\def\rank{{\rm rank}}

\def\Diff{{\rm Diff}}

\def\ind{{\rm ind}}
\def\Jac{{\rm Jac}}

\def\End{{\rm End}}
\def\Tr{{\rm Tr}}
\def\Id{{\rm Id}}
\def\vol{{\rm vol}}

\def\p{\partial}

\def\ch{{\rm ch}}

\def\Lie{{\rm Lie}}

\def\Si{{\Sigma}}

\def\inbar{\,\vrule height1.5ex width.4pt depth0pt}
\def\r{{\rm Re}}
\def\i{{\rm Im}}

\def\a{\alpha}
\def\b{\beta}
\def\g{\gamma}
\def\d{\delta}
\def\e{\epsilon}

\def\la{\lambda}
\def\th{\theta}

\def\om{\omega}

\def\bar{\overline}
\def\z{\zeta}
\def\om{\omega}

\def\Si{{\Sigma_{g}}}
\def\G{\Gamma}

\def\det{{\rm det}}
\def\tr{{\rm tr}}

\def\Tr{{\rm Tr}}

\def\IH{{\bf H}}

\chardef\tempcat=\the\catcode`\@
\catcode`\@=11
\def\cyracc{\def\u##1{\if \i##1\accent"24 i
    \else \accent"24 ##1\fi }}
\newfam\cyrfam
\font\tencyr=wncyr10
\def\cyr{\fam\cyrfam\tencyr\cyracc}


\lref\HH{
  J.~B.~Hartle and S.~W.~Hawking,
  ``Wave Function Of The Universe,''
  Phys.\ Rev.\ D {\bf 28}, 2960 (1983).}

\lref\ovv{H.~Ooguri, C.~Vafa and E.~Verlinde, ``Hartle-Hawking wave-function for
flux compactifications: the entropic principle,''
hep-th/0502211.}

\lref\osv{H.~Ooguri, A.~Strominger and C.~Vafa, ``Black hole attractors and the
topological string,'' Phys.\ Rev.\ D
{\bf 70}, 106007 (2004) [arXiv:hep-th/0405146].}

\lref\cano{ P.~Candelas and X.~de la Ossa,
  ``Moduli Space Of Calabi-Yau Manifolds,''
  Nucl.\ Phys.\ B {\bf 355}, 455 (1991).}

\lref\attra{ S.~Ferrara, R.~Kallosh and A.~Strominger,
  ``N=2 extremal black holes,''
  Phys.\ Rev.\ D {\bf 52}, 5412 (1995).
}


\lref\RD{
R.~Dijkgraaf, ``A Universal Wave Function,''
talk at the ``{\it Black holes, topological strings,
and invariants of holomorphic submanifolds}'',
Harvard University, January 2006.
}

\lref\Gtalks{
A.~Gerasimov,  ITEP seminars, 2000-2001.
}

\lref\RDlec{
  R.~Dijkgraaf,
  ``Fields, strings and duality,''
  arXiv:hep-th/9703136.
  }

\lref\Mor{
  A.~Morozov,
  ``Identities between quantum field theories in different dimensions,''
  arXiv:hep-th/9810031.
  }

\lref\GM{
G.~Moore, ``Uncertainty and Fluxes,''
talk at the  ``{\it Black holes, topological strings,
and invariants of holomorphic submanifolds}'',
Harvard University,  January 2006.
}

\lref\Hgeom{
  N.~J.~Hitchin,
  ``The geometry of three-forms \ in six and seven   dimensions,''
  arXiv:math.dg/0010054.
}

\lref\Hstable{
  N.~J.~Hitchin,
  ``Stable forms and special metrics,''
  arXiv:math.dg/0107101.
}

\lref\TM{
  R.~Dijkgraaf, S.~Gukov, A.~Neitzke and C.~Vafa,
  ``Topological M-theory as unification of form theories of gravity,''
  arXiv:hep-th/0411073.
}

\lref\NOV{
  N.~Nekrasov, H.~Ooguri and C.~Vafa,
  ``S-duality and topological strings,''
  JHEP {\bf 0410}, 009 (2004)
  arXiv:hep-th/0403167.
}

\lref\Semyonych{A.~S.~Losev, ``Perspectives of string theory,''
talk at the ``{\it String theory at Greater Paris}''
seminar, 2001, as cited in \NOV.
}

\lref\Nikita{
  N.~Nekrasov,
  ``A la recherche de la m-theorie perdue. Z theory: Chasing m/f theory,''
  arXiv:hep-th/0412021.
}

\lref\Nikitac{
  A.~Losev, G.~W.~Moore, N.~Nekrasov and S.~Shatashvili,
  ``Chiral Lagrangians, anomalies, supersymmetry, and holomorphy,''
  Nucl.\ Phys.\ B {\bf 484}, 196 (1997),
  arXiv:hep-th/9606082.
}

\lref\GS{
  A.~A.~Gerasimov and S.~L.~Shatashvili,
  ``Towards integrability of topological strings.~I: Three-forms on Calabi-Yau
  manifolds,''
  JHEP {\bf 0411}, 074 (2004),
  arXiv:hep-th/0409238}

\lref\Grassi{
  P.~A.~Grassi and P.~Vanhove,
  ``Topological M theory from pure spinor formalism,''
  arXiv:hep-th/0411167.
  }

\lref\Boer{
  J.~de Boer, A.~Naqvi and A.~Shomer,
  ``The topological G(2) string,''
  arXiv:hep-th/0506211.
  }

\lref\Anguelova{
  L.~Anguelova, P.~de Medeiros and A.~Sinkovics,
  ``Topological membrane theory from Mathai-Quillen formalism,''
  arXiv:hep-th/0507089.
}

\lref\BTZ{
  G.~Bonelli, A.~Tanzini and M.~Zabzine,
  ``On topological M-theory,''
  arXiv:hep-th/0509175.
  }

\lref\BTtop{
  L.~Baulieu and A.~Tanzini,
  ``Topological symmetry of forms, N = 1 supersymmetry and S-duality on special
  manifolds,''
  arXiv:hep-th/0412014.
  }

\lref\Calabi{
  E.~Calabi, ``An intrinsic characterization of harmonic 1-forms'',
  Global Analysis, Papers in Honor of K.Kodaira, (D.C.Spencer and S.Iyanaga, ed.),
  1969, pp. 101-107.
  }

\lref\Novikov{
  S.~P.~Novikov, ``Topology of Foliations given by the
   real part of holomorphic one-form,''  arXiv: math.GT/0501338.
   }

\lref\KZ{
  M.~Kontsevich and A.~Zorich,
  ``Lyapunov exponents and Hodge theory,''
  arXiv:hep-th/9701164.
}

\lref\GSV{
  S.~Gukov, K.~Saraikin and C.~Vafa,
  ``The entropic principle and asymptotic freedom,''
  arXiv:hep-th/0509109.
  }

\lref\ACGH{
  E.~Arbarello, M.~Cornalba, P.A. Griffiths and J. Harris, ``Geometry of Algebraic
  Curves,'' Vol.1. Berlin, Heidelberg, New-York and Tokyo, 1985.
}

\lref\FK{
  H.~M.~Farkas and I.~Kra, ``Riemann Surfaces,'' Graduate Texts in Mathematics, no. 71,
  2nd ed., Springer Verlag, Berlin and New York, 1991.
 }

\lref\Jost{
J.~Jost, ``Compact Riemann Surfaces,''  Springer-Verlag and New York, 2002.
}

\lref\GK{
  K.~Gawedzki and A.~Kupiainen,
  ``Coset Construction From Functional Integrals,''
  Nucl.\ Phys.\ B {\bf 320}, 625 (1989).
  }

\lref\Gaw{
  K.~Gawedzki,
  ``Quadrature Of Conformal Field Theories,''
  Nucl.\ Phys.\ B {\bf 328}, 733 (1989).
  }

\lref\Sglas{
  M.~Spiegelglas and S.~Yankielowicz,
  ``G / G topological field theories by cosetting G(k),''
  Nucl.\ Phys.\ B {\bf 393}, 301 (1993)
  arXiv:hep-th/9201036.
  }

\lref\WitHol{
  E.~Witten,
  ``On Holomorphic factorization of WZW and coset models,''
  Commun.\ Math.\ Phys.\  {\bf 144}, 189 (1992).
}

\lref\BT{
  M.~Blau and G.~Thompson,
  ``Derivation of the Verlinde formula from Chern-Simons theory and the G/G
  model,''
  Nucl.\ Phys.\ B {\bf 408}, 345 (1993)
  arXiv:hep-th/9305010.
}

\lref\Ger{
  A.~Gerasimov,
  ``Localization in GWZW and Verlinde formula,''
  arXiv:hep-th/9305090.
  }

\lref\WitGrass{
  E.~Witten,
  ``The Verlinde algebra and the cohomology of the Grassmannian,''
  arXiv:hep-th/9312104.
}

\lref\WitJon{
  E.~Witten,
  ``Quantum Field Theory And The Jones Polynomial,''
  Commun.\ Math.\ Phys.\  {\bf 121}, 351 (1989).
  }

\lref\MS{
G.~Moore and N.~Seiberg,
``Lectures on RCFT,'' in {\it Superstrings '89: Proceedings of the
Trieste Spring School}, pp. 1--129, Ed. by M.~Green et al, World
Scientific, Singapore, 1990.
}

\lref\APW{
  S.~Axelrod, S.~Della Pietra and E.~Witten,
  ``Geometric Quantization Of Chern-Simons Gauge Theory,''
  J.\ Diff.\ Geom.\  {\bf 33}, 787 (1991).
}

\lref\Moore{G.~Moore, ``Arithmetic and attractors,'' arXiv:hep-th/9807087.}

\lref\GV{S.~Gukov and C.~Vafa, ``Rational conformal field theories and complex
multiplication,'' Commun.\ Math.\ Phys.\
{\bf 246}, 181 (2004), hep-th/0203213.
}

\lref\Chen{M.~Chen, ``Complex multiplication, rationality and
mirror symmetry for abelian varieties'' math.AG/0512470.
}

\lref\EMSS{
  S.~Elitzur, G.~W.~Moore, A.~Schwimmer and N.~Seiberg,
  ``Remarks On The Canonical Quantization Of The Chern-Simons-Witten Theory,''
  Nucl.\ Phys.\ B {\bf 326}, 108 (1989).
  }

\lref\MSzoo{
  G.~W.~Moore and N.~Seiberg,
  ``Taming The Conformal Zoo,''
  Phys.\ Lett.\ B {\bf 220}, 422 (1989).
 }

\lref\BN{
  M.~Bos and V.~P.~Nair,
  ``Coherent State Quantization Of Chern-Simons Theory,''
  Int.\ J.\ Mod.\ Phys.\ A {\bf 5}, 959 (1990).
}

\lref\ih{
  M.~Aganagic, R.~Dijkgraaf, A.~Klemm, M.~Marino and C.~Vafa,
  ``Topological strings and integrable hierarchies,''
  Commun.\ Math.\ Phys.\  {\bf 261}, 451 (2006)
  arXiv:hep-th/0312085.
  }

\lref\qym{
  M.~Aganagic, H.~Ooguri, N.~Saulina and C.~Vafa,
  ``Black holes, q-deformed 2d Yang-Mills, and non-perturbative topological
  strings,''
  Nucl.\ Phys.\ B {\bf 715}, 304 (2005)
  arXiv:hep-th/0411280.
  }

\lref\PW{
  V.~Pestun and E.~Witten,
  ``The Hitchin functionals and the topological B-model at one loop,''
  Lett.\ Math.\ Phys.\  {\bf 74}, 21 (2005)
  arXiv:hep-th/0503083.
}
  \lref\Pestun{
  V.~Pestun,
  ``Black hole entropy and topological strings on generalized CY manifolds,''
  arXiv:hep-th/0512189.
  }

\lref\Vafa{
C.~Vafa, private communication
}

\lref\Vswamp{
  C.~Vafa,
  ``The string landscape and the swampland,''
  arXiv:hep-th/0509212.
  }

\lref\Coleman{R.~Coleman, ``Torsion Points on Curves,''
In Galois representations and arithmetic algebraic geometry
(Y.~Ihara ed.), Adv. Studies Pure Math. {\bf 12} (1987) 235.}

\lref\GMMS{
  S.~Gukov, E.~Martinec, G.~W.~Moore and A.~Strominger,
  ``Chern-Simons gauge theory and the AdS(3)/CFT(2) correspondence,''
}

\lref\BM{
  D.~Belov and G.~W.~Moore,
  ``Classification of abelian spin Chern-Simons theories,''
  arXiv:hep-th/0505235.
}

\lref\BPZ{
  A.~A.~Belavin, A.~M.~Polyakov and A.~B.~Zamolodchikov,
  ``Infinite Conformal Symmetry In Two-Dimensional Quantum Field Theory,''
  Nucl.\ Phys.\ B {\bf 241}, 333 (1984).
  }

\lref \Polyakov{
Polyakov, 2006
}

\lref\Verlinde{
  E.~P.~Verlinde,
  ``Global aspects of electric - magnetic duality,''
  Nucl.\ Phys.\ B {\bf 455}, 211 (1995)
  arXiv:hep-th/9506011
  }

\lref \HHer{
  S.~W.~Hawking and T.~Hertog,
  ``Populating the Landscape: A Top Down Approach,''
  arXiv:hep-th/0602091.
}

\lref\WitTop{
  E.~Witten,
  ``AdS/CFT correspondence and topological field theory,''
  JHEP {\bf 9812}, 012 (1998)
  arXiv:hep-th/9812012.
  }

\lref\MMM{
  A.~Marshakov, A.~Mironov and A.~Morozov,
  ``On equivalence of topological and quantum 2-d gravity,''
  Phys.\ Lett.\ B {\bf 274}, 280 (1992)
  arXiv:hep-th/9201011.
  }

\lref\DVV{
  R.~Dijkgraaf, H.~L.~Verlinde and E.~P.~Verlinde,
  ``Loop Equations And Virasoro Constraints In Nonperturbative 2-D Quantum
  Gravity,''
  Nucl.\ Phys.\ B {\bf 348}, 435 (1991).
  }

\lref\BCOV{
  M.~Bershadsky, S.~Cecotti, H.~Ooguri and C.~Vafa,
  ``Kodaira-Spencer theory of gravity and exact results for quantum string
  amplitudes,''
  Commun.\ Math.\ Phys.\  {\bf 165}, 311 (1994),
  arXiv:hep-th/9309140.
  }

\lref\DVafaV{
  R.~Dijkgraaf, C.~Vafa and E.~Verlinde,
  ``M-theory and a Topological String Duality,''
  arXiv:hep-th/0602087.
  }

\lref\DMW{
  D.~E.~Diaconescu, G.~W.~Moore and E.~Witten,
  ``E(8) gauge theory, and a derivation of K-theory from M-theory,''
  Adv.\ Theor.\ Math.\ Phys.\  {\bf 6}, 1031 (2003),
  arXiv:hep-th/0005090.
}

\lref\Vop{
  C.~Vafa,
  ``Operator Formulation On Riemann Surfaces,''
  Phys.\ Lett.\ B {\bf 190}, 47 (1987).
}

\lref\Tor{
  A.~Giveon, M.~Porrati and E.~Rabinovici,
  ``Target space duality in string theory,''
  Phys.\ Rept.\  {\bf 244}, 77 (1994),
  arXiv:hep-th/9401139.
}

\lref\Wittgrav{
  E.~Witten,
  ``On The Structure Of The Topological Phase Of Two-Dimensional Gravity,''
  Nucl.\ Phys.\ B {\bf 340}, 281 (1990).
}

\lref\Verfor{
  E.~P.~Verlinde,
  ``Fusion Rules And Modular Transformations In 2-D Conformal Field Theory,''
  Nucl.\ Phys.\ B {\bf 300}, 360 (1988).
}

\lref\Nikitafive{
  N.~Nekrasov,
  ``Five dimensional gauge theories and relativistic integrable systems,''
  Nucl.\ Phys.\ B {\bf 531}, 323 (1998),
  arXiv:hep-th/9609219.
  }

\lref\LN{
  A.~E.~Lawrence and N.~Nekrasov,
  ``Instanton sums and five-dimensional gauge theories,''
  Nucl.\ Phys.\ B {\bf 513}, 239 (1998)
  arXiv:hep-th/9706025.
  }

\lref\WitNM{
  E.~Witten,
  ``The N matrix model and gauged WZW models,''
  Nucl.\ Phys.\ B {\bf 371}, 191 (1992).
}

\lref\HitGen{N.~Hitchin,
 "Generalized Calabi-Yau manifolds",
 Quart.\ J.\ Math.\ Oxford\ Ser. {\bf 54}, 281 (2003),
    arXiv:math.DG/0209099.
}

\lref\Gualtieri{
M.~Gualtieri,
``Generalized complex geometry,''
Oxford University DPhil thesis, arXiv:math.DG/0401221.
}

\lref\BTKS{
  M.~Blau, F.~Hussain and G.~Thompson,
  ``Some General Aspects of Coset Models and Topological Kazama-Suzuki
  Models,''
  Nucl.\ Phys.\ B {\bf 488}, 541 (1997),
  arXiv:hep-th/9510187.
 }

\lref\FKS{
  S.~Ferrara, R.~Kallosh and A.~Strominger,
  ``N=2 extremal black holes,''
  Phys.\ Rev.\ D {\bf 52}, 5412 (1995),
  arXiv:hep-th/9508072.
}

\lref\Strominger{
  A.~Strominger,
  ``Macroscopic Entropy of $N=2$ Extremal Black Holes,''
  Phys.\ Lett.\ B {\bf 383}, 39 (1996)
  arXiv:hep-th/9602111.
}

\lref\CWM{
  G.~Lopes Cardoso, B.~de Wit and T.~Mohaupt,
  ``Corrections to macroscopic supersymmetric black-hole entropy,''
  Phys.\ Lett.\ B {\bf 451}, 309 (1999),
  arXiv:hep-th/9812082.
}

\lref\Kiritsis{
  E.~B.~Kiritsis,
  ``Duality in gauged WZW models,''
  Mod.\ Phys.\ Lett.\ A {\bf 6}, 2871 (1991).
}

\lref\GSVstr{
  S.~Gukov, K.~Saraikin and C.~Vafa,
  ``A stringy wave function for an S**3 cosmology,''
  arXiv:hep-th/0505204.
  }

\lref\Sar{
K.~Saraikin, "On holomorphic factorization
of the topological string partition function",
to appear.
}

\lref\WitEff{
  E.~Witten,
  ``Duality relations among topological effects in string theory,''
  JHEP {\bf 0005}, 031 (2000),
  arXiv:hep-th/9912086.
  }

 \lref\Nikitabg{
  N.~Nekrasov,
  ``Lectures on curved beta-gamma systems, pure spinors, and anomalies,''
  arXiv:hep-th/0511008.
}

\lref\AGMV{
  L.~Alvarez-Gaume, G.~W.~Moore and C.~Vafa,
  ``Theta Functions, Modular Invariance, And Strings,''
  Commun.\ Math.\ Phys.\  {\bf 106}, 1 (1986).
}

\lref\VVc{
  E.~P.~Verlinde and H.~L.~Verlinde,
  ``Chiral Bosonization, Determinants And The String Partition Function,''
  Nucl.\ Phys.\ B {\bf 288}, 357 (1987).
  }

\lref\DVVc{
  R.~Dijkgraaf, E.~P.~Verlinde and H.~L.~Verlinde,
  ``C = 1 Conformal Field Theories On Riemann Surfaces,''
  Commun.\ Math.\ Phys.\  {\bf 115}, 649 (1988).
  }

\lref\Tyurin{A.~N.~Tyurin,
"Quantization, Classical and Quantum Field Theory and Theta Functions",
CRM monograph series, vol. 32, American Mathematical Society, Providence, RI, 2003,
math.AG/0210466.
}

\lref\Polyakovb{
  A.~M.~Polyakov,
  ``Quantum Geometry Of Bosonic Strings,''
  Phys.\ Lett.\ B {\bf 103}, 207 (1981).
}

\lref\IM{
  C.~Imbimbo and S.~Mukhi,
  ``The Topological matrix model of c = 1 string,''
  Nucl.\ Phys.\ B {\bf 449}, 553 (1995),
  arXiv:hep-th/9505127.
}

\lref\BK{
  A.~A.~Belavin and V.~G.~Knizhnik,
  ``Algebraic Geometry And The Geometry Of Quantum Strings,''
  Phys.\ Lett.\ B {\bf 168}, 201 (1986).
}

\lref\Witbckrnd{
  E.~Witten,
  ``Quantum background independence in string theory,''
  arXiv:hep-th/9306122.
}


\Title{\vbox{\baselineskip11pt
\hbox{hep-th/0604176}
\hbox{HUTP-06/A0014}
\hbox{ITEP-TH-16/06}
}}
{\vbox{
\centerline{Abelian Varieties,  RCFTs, Attractors, }
\medskip
\centerline{and Hitchin Functional in Two Dimensions }
}}
\centerline{
Kirill Saraikin\footnote{$^{\dagger}$}{\cyr Komandirovan iz
ITF im. L.D.Landau, 119334 Moskva, ul. Kosygina 2,  i
ITE1F, 117259 Moskva, ul. B.Cheremushkinskaya 25, Rossiya.}}
\medskip
\vskip 8pt
\centerline{ \it Jefferson Physical Laboratory, Harvard University,}
\centerline{ \it Cambridge, MA 02138, USA}
\vskip 30pt
\noindent
We consider a generating function
for the number of conformal blocks
in  rational conformal field theories with an
even central charge $c$ on a genus $g$ Riemann surface.
It defines an entropy functional on the moduli space of
conformal field theories and is captured by the
 gauged WZW model whose target space
 is an abelian variety. We study a  special coupling of
this theory to
two-dimensional gravity.
When $c=2g$, the coupling is
non-trivial due to the gravitational instantons,
and the action of the theory can be interpreted as
a two-dimensional analog of the Hitchin functional for
Calabi-Yau manifolds.
This gives rise to the effective action on the moduli
space of Riemann surfaces, whose  critical points
are attractive and correspond to Jacobian varieties
admitting complex multiplication.
The theory that we describe can be viewed as a
dimensional reduction of  topological M-theory.

\medskip
\Date{April, 2006}

\listtoc\writetoc

\medskip
\medskip

\newsec{Introduction}

The purpose of this paper is to look at a
two-dimensional toy model of the topological M-theory \refs{\TM, \Nikita}
in order to gain some insights on its plausible
quantum description.
The analysis of this model supports the idea that
quantum partition function of the topological M-theory
is given by a generalized index theorem for the
moduli space.
In particular, this implies that the
OSV conjecture  \osv\ should be viewed  as a higher
dimensional analog of the E.~Verlinde's formula for the number of
conformal blocks
in a two-dimensional conformal field theory.
Below we briefly sketch  some relatively old  ideas that provide a
motivation for this picture\foot{This introductory section is inspired
by the talks of R.~Dijkgraaf \RD\ and A.~Gerasimov~\Gtalks.}.

\subsec{Universal Partition Function and Universal Index Theorem}

It is well known that many generic features of
topological  theories can be nicely
described using the category theory.
Roughly speaking, the translation between the
category theory and quantum mechanics language goes as follows
(for more details and references see, $e.g.$ \RDlec).
One starts with associating a wave-function $| \Psi_0 (M)\rangle$ to
a $d$-dimensional manifold $M$:
\eqn\wf{
{\lower22.0pt \hbox{\epsfxsize0.28in\epsfbox{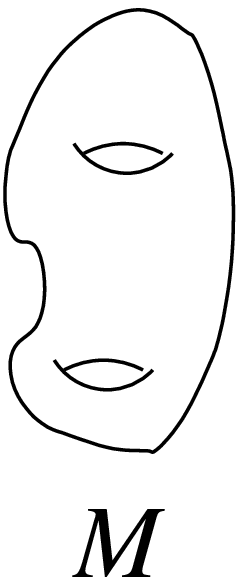}}}  \ =
\ | \Psi_0 (M)\rangle.
}
A natural generalization
is assigning some additional
structures $E$
(bundles, sheaves, gerbes,~etc.) to $M$:
$ | \Psi_0 (M)\rangle \to  | \Psi_E (M)\rangle$.
In the  language of physics, this is equivalent to
putting some branes and/or fluxes on $M$.
The morphisms in the category of $(d+1)$-dimensional
manifolds with extra structures are bordisms $E \to  F$:
\eqn\prop{
{\lower22.0pt \hbox{\epsfxsize1.2in\epsfbox{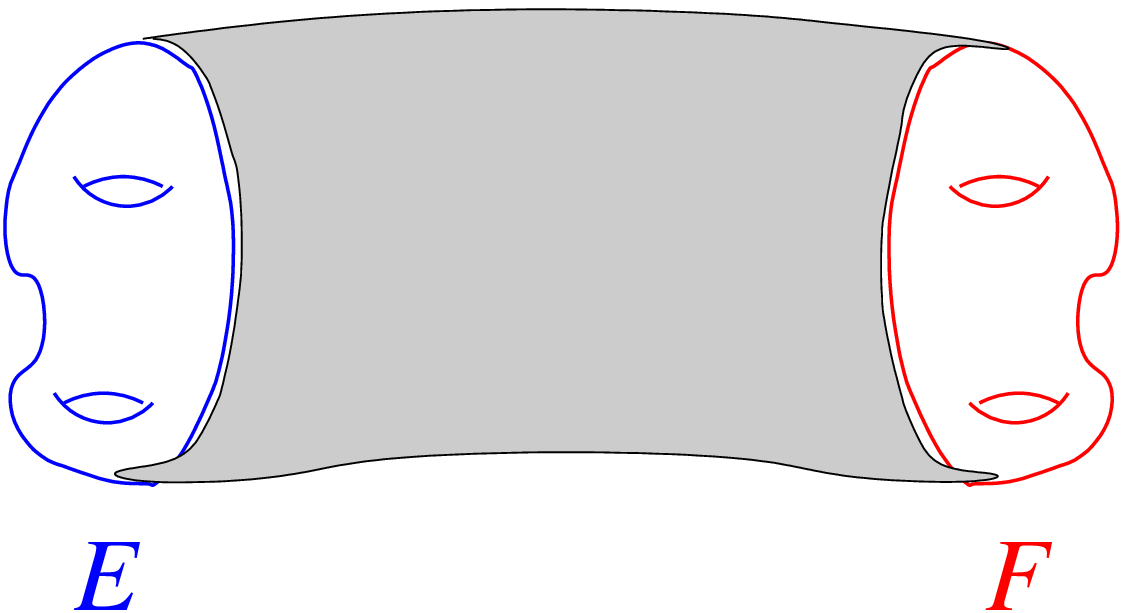}}} \  = \
\langle\Psi_E (M) | \Psi_F (M)\rangle,
}
interpreted as quantum mechanical  propagators
between the states $E$ and $F$.
The composition law of two bordisms given by "gluing"
two boundaries
is a basic feature of the functional integral:
\eqn\pprop{
\langle\Psi_E | \Psi_G \rangle =
\int \CD F \langle\Psi_E  | \Psi_F \rangle \langle\Psi_F  | \Psi_G\rangle.
}
These pictures, of course,
mimic the well-known operations with the world-sheets in  string theory.
The universal partition function $Z_{M\times \S^1}$ is assigned to the
manifold of the form $M\times \S^1$:
\eqn\trc{
{\lower31.0pt \hbox{\epsfxsize1.1in\epsfbox{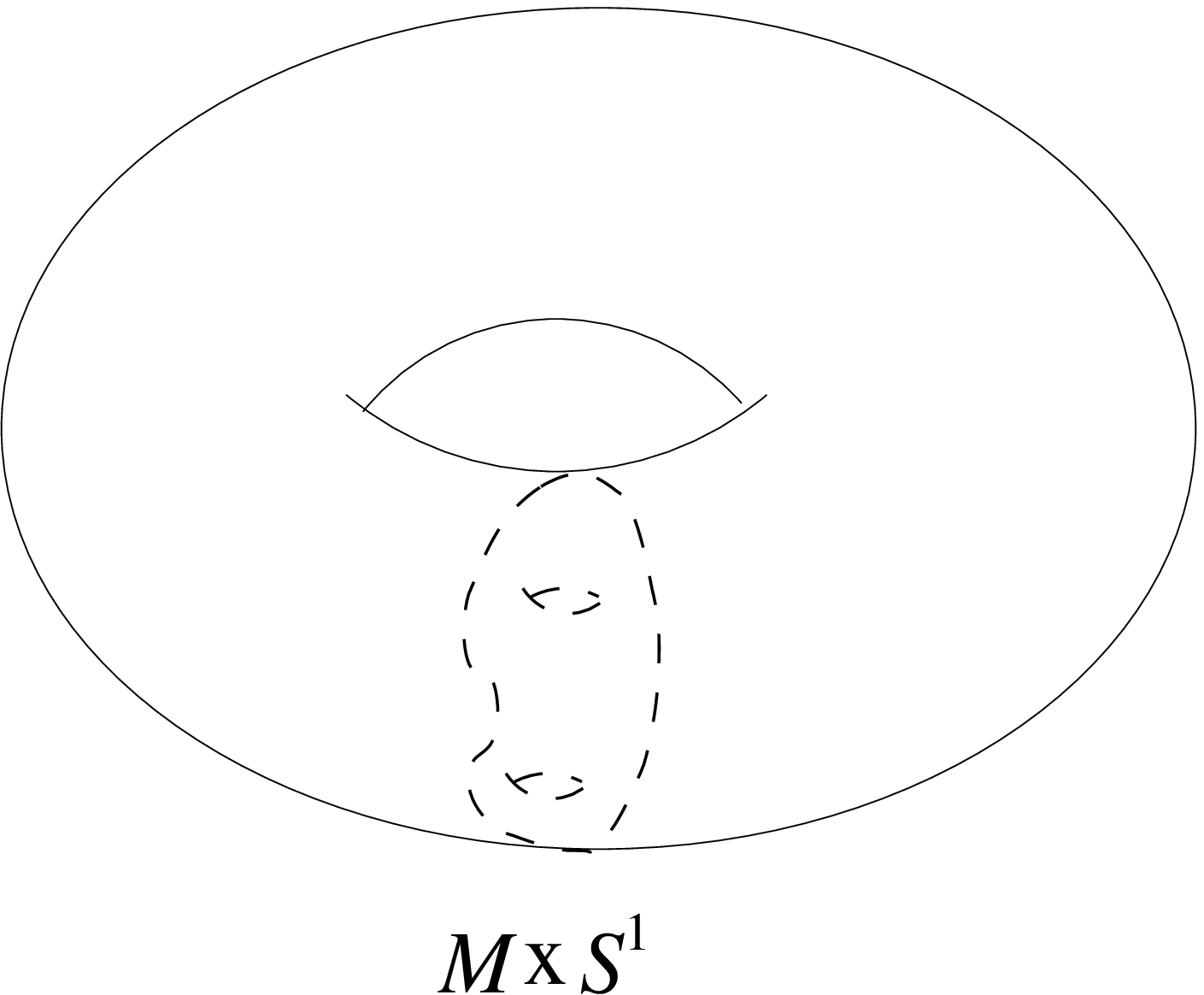}}} \ =\
\Tr_{E} \langle\Psi_E (M)| \Psi_E (M)\rangle.
}
Roughly speaking, it counts the number of
topological (massless) degrees
of freedom, or (super) dimension of the corresponding Hilbert space.
Relation \trc\ is  a
manifestation of the  equivalence between the
Lagrangian and Hamiltonian formulation
of the path integral.
In the framework of geometric quantization,
the Hilbert space is given by the
cohomology groups of  the moduli space $\CM_E$ of
$E$-structures on $M$, with coefficients
in the (prequantum) line bundle $\CL$.
Therefore, the universal partition function is
associated with the  corresponding index:
$Z_{M\times \S^1} = { \rm Ind \CD_E}$, where\foot{
In a more general setup $ \dim H^{\bullet}$
is substituted by $  \Tr_{H^{\bullet}} \CD_E  $.}
\eqn\inddef{
{ \rm Ind \CD_E} =
\sum_{n} (-1)^n \dim H^{n} (\CM_E, \CL).
}
In many interesting cases  higher cohomology groups vanish, and
the partition function computes
the dimension of the Hilbert space:
$\dim Hilb_{\CM_E} = \dim  H^{0} (\CM_E, \CL)$.
On the other hand, the partition function can also be computed
via the universal index theorem:
\eqn\indth{
{ \rm Ind \CD_E}  = \int_{\CM_E}
 \ch ( \CL) {\rm Td} (T \CM_E),
}
where the integral over the moduli space
arises after localization in the functional integral\foot{This
formula is very schematic, and its exact form depends on the details
of the problem. For example, twisting by $K^{1/2}$ will
result in appearance of $\hat A$ instead of the Todd class.}.

Moreover, since one can think of the wave-function \wf\
as of a partition function itself:
$Z_M =|\Psi(M)\rangle $,
the definitions \prop-\trc\ imply
the quadratic relation of the form
\eqn\qr{
Z_{M\times \S^1} \sim Z_M Z_M^*.
}
If $M$ is a (generalized)
complex manifold, this can be  true even at the level
of the Lagrangian for the local (massive) degrees
of freedom. Indeed,
in the functional integral formalism
we are dealing with the
generalized Laplacian operator $\Delta_E =  \CD^{\dag}_E \CD_E$
constructed from the generalized Dirac operator $\CD_E$.
The square factor for the local degrees of freedom
arises  from the Quillen theorem:
\eqn\qth{
\det' \Delta_E = e^{-  \CA(\CD_E) } \big| \det' \CD_E \big|^2.
}
Here $\CA(\CD_E)$ is the holomorphic anomaly:
$\p \bar \p \CA(\CD_E) \not= 0$.
It is natural to assign this anomaly to the integration
measure over the moduli space, and then interpret the deviation
from the quadratic relation \qr\ as a quantum correction.

Below we list some examples  that illustrate these phenomena,
which sometimes is referred to  as the bulk/boundary
correspondence (for more examples, see,
$e.g.$, \refs{\TM, \Nikita, \WitTop, \Mor}).
\vskip 0.8cm
\vbox{
\centerline{\vbox{
\hbox{\vbox{\offinterlineskip
\def\tablespace{height7pt&\omit&&\omit&&\omit&\cr}
\def\tablerule{\tablespace\noalign{\hrule}\tablespace}

\hrule\halign{&\vrule#&\strut\hskip0.2cm\hfill
#\hfill\hskip0.2cm\cr
\tablespace & Dimension && Correspondence && Index Theorem  &\cr
\tablerule & $5_{\IC}+1$  && M-theory/Type IIA   &&
$Z_{\rm M} \sim Z^L_{\rm IIA}  Z^R_{\rm IIA}$ &\cr
\tablerule & $4_{\IC}+1$  && ?/8$d$ Donaldson-like theory &&  ? &\cr
\tablerule & $3_{\IC}+1$  &&$G_2$/CY$_{3}$ in topological M-theory  &&
$Z_{\rm BH} \sim |Z_{\rm top}|^2 $  &\cr
\tablerule & $2_{\IC}+1$  && $5d$ SYM/Donaldson-Witten theory &&
$Z_{\rm SYM} \sim Z_{\rm DW}Z_{\rm DW}^* $ &\cr
\tablerule & $1_{\IC}+1$  && Chern-Simons theory/CFT && $Z_{\rm CS}
\sim |Z_{\rm CFT}|^2$  &\cr
\tablerule & $0_{\IC}+1$  &&$2d$ quantum gravity/$2d$ topological gravity &&
$Z_{\rm qg} \sim Z_{\rm tg}^2$ &\cr
\tablespace}\hrule}}}}
\centerline{ \hbox{{\bf Table 1:}{\it ~~ Examples of the bulk/boundary
correspondence.}}}
}
\vskip 0.5cm

Relation of the type \qr\  is known
in the context of
the matrix models ($0_{\IC}+1$ dimensions)
as a manifestation of the correspondence between
the quantum gravity and topological gravity in two dimensions
(see, $e.g.$ \refs{\Wittgrav, \MMM}).
Another form of this relation  is  $\tau = \sqrt{Z_{qg}} $,
where $\tau$ is a tau-function of the KdV hierarchy~\DVV.
The index formula \indth\   represents computation
of the Euler characteristic of $\CM_{g,h}$ via the
Penner matrix model.

In $1_{\IC}+1$ dimensions \qr\
is the famous relation between the Chern-Simons theory and two-dimensional
conformal field theory \WitJon.
The index theorem in this case
gives  the E.~Verlinde's formula for the
number of conformal blocks \Verfor.
It will be the subject of primary interest of this paper.

In $2_{\IC}+1$ dimensions \indth\ and \qr\
express  computation of the Gromov-Witten invariants from
the counting of the  BPS states in
the five dimensional
supersymmetric gauge theory
\refs{\Nikitafive, \LN}.

In $3_{\IC}+1$  dimensions relation \qr\ is known
as the OSV conjecture \osv.
The exact formulation of the index theorem \indth\ in this
case is not known, and the question about the non-pertubative (quantum)
corrections to \qr\ is very important
for clarifying the relation between the topological
strings and the black holes entropy.
It is expected that the answer
can be given  in the framework of the
topological M-theory  \TM\ (which was called
Z-theory in \Nikita, see also
\refs{\GS,  \NOV, \Grassi, \BTtop, \Boer, \Anguelova, \BTZ,\DVafaV} for
a discussion on the related issues).

Not much is known about the $4_{\IC}+1$ dimensional
example,  apart from its
relation to the Donaldson-like theory in eight dimensions \Nikita.

The M-theory/Type IIA relation (topological
version of which is the $5_{\IC}+1$ dimensional example)
was a source of tremendous progress
in string theory over the
last decade. Needless to say,
there are many subtle details
involved in this correspondence
(see, $e.g.$~\DMW).

The $5_{\IC}+1$ dimensions is not the end of the story,
it  probably continues to higher dimensions (F-theory, etc.).
Also, it is worth mentioning
that there are many signs for the (hidden) integrability in these
theories, which is intimately
related to the free fermion representation.
It allows for a tau-function
interpretation of the partition function
and is responsible for the appearance of the
integrable hierarchies.

Finally, let us note that the theories
in different dimensions from Table 1 are connected
(apart from the obvious dimensional reduction)
via the generalized transgression and descent equations
(at least at the classical level), which allow
for  going one complex dimension up or down.
From the geometry-categoric  viewpoint
this is related to the sequence of
 $d$-manifolds serving as a boundaries for $d+1$-manifolds:
$$0 \to M_1 \to \ldots \to M_d \to M_{d+1} \to \ldots$$
By analogy with \Nikita, one can call the theory unifying these theories in
different dimensions the Z-theory.

\subsec{ The Entropic Principle
and Quantum Mechanics on the Moduli Space}

So far we discussed   theories describing
topological invariants of some structures living
on a fixed $d$-manifold $M$. It is interesting to ask how
these invariants, captured by the universal
partition function \trc, change if we vary $M$
within its topological class.
For example, we can talk about transport on the
moduli space of genus $g$ Riemann surfaces $\CM_{g}$ or
even fantasize about the moduli space of
"all" Calabi-Yau threefolds $\CM_{CY_3}$.

This question typically arise in quantum gravity,
where one is interested in comparing different possible
universes (vacuum states) $M$ and choosing "preferred" ones.
The original suggestion of Hartle and Hawking \HH\ is
to weight each vacuum by the probability of creation from
nothing (see also recent discussion in \HHer).
This gives some measure on the landscape of vacua.
In \ovv, this proposal was interpreted
in the context of string   compactification with fluxes
on $\AdS_2\times\S^2\times M$, where $M$ is a
 Calabi-Yau threefold, using
the OSV conjecture \osv. The weight, associated to
a given $M$, is the
norm of the Hartle-Hawking wave-function, which
is related to the  entropy $S_{p,q}$ of the dual black hole,
obtained by wrapping a $D3$ brane with magnetic and electric charges $(p,q)$
on $M$. This is called the entropic principle \refs{\ovv, \GSV}:
\eqn\entrop{
\langle\Psi_{p,q}(M)|\Psi_{p,q}(M) \rangle \sim \exp ({S_{p,q}})
}
The complex moduli of $M$ are fixed by the charges
$(p,q)$ via the attractor mechanism
\refs{\FKS,\Strominger,\CWM,\Moore}.

The entropic principle implies that one can
define corresponding
quantum mechanical problem on the moduli space $\CM_M$
by summing over  all Calabi-Yau manifolds $M$ with the weight \entrop.
This path integral can be used, for example,
 for computing  correlation functions of the gravitation fluctuations
around the points on the moduli space
that correspond to "preferred"  string   compactifications
(see \GSVstr\ for some steps in this direction).
This approach can help us
to  shed some light on
the fundamental physical problems, such as
quantum cosmology and string landscape.

The entropic principle in general can be formulated by saying that
the  entropy function is the Euler characteristic of the moduli space,
associated with the problem.
It is expected that critical points of the
entropy function on the moduli space correspond to
special manifolds with extra (arithmetic) structures,
such as complex multiplication, Lie algebra lattices, etc.
There is also a hidden integrality involved
coming from the quantization of  (at least partially)
compact moduli space.
As a result, we expect appearance of the nice modular functions and
automorphic forms at the critical points of the
entropy function.
This will be very clear in a simple two-dimensional
example considered in the paper.

It was noted in \refs{\TM,\Nikita, \GS} that some new
geometric functionals introduced by Hitchin \refs{\Hgeom, \Hstable}
might be useful for  formulation
of this problem
in the context of topological strings.
There are several reasons why the approach based on the Hitchin
functional is attractive.
Since it is a diffeomorphism-invariant functional
depending only on the
cohomology classes of some differential forms,
it is a proper candidate  for the description of  topological
degrees of freedom. It can also be used for
incorporating the generalized geometry moduli.
Moreover, at the classical level it reproduces
the black hole entropy.
We suggest that in order to  use the Hitchin
functional for a quantum mechanical description
of the moduli space one has to
study   variation of the
cohomology classes that are usually fixed.
The space of cohomologies has a natural
symplectic structure, defined by the cup product,
and therefore is easy to quantize.
Moreover, the mapping class   group $U={\Diff /\Diff_0}$ acts
naturally on  the space of cohomologies,
which might be  useful for the non-perturbative
description of the theory.

In this paper we take a modest step
in this direction by applying this idea to the
two-dimensional toy model, that has many interesting
features which are expected to survive in higher dimensions.
The advantage of taking digression to the
two dimensions is that in this case
(almost) everything becomes solvable.
Our goal is to find a two-dimensional sibling
of the Hitchin functional,
formulate an analog of the entropic principle in $1_{\IC}+1$
dimensions, and describe corresponding quantum theory.
It turns out that this way we  find  a
unified description of all two-dimensional topologies.
Moreover, study of  the two-dimensional  model leads to a
natural generalization of the six-dimensional Hitchin
functional, which may be  useful for    understanding
  of the topological M-theory at quantum level \Sar.

\bigskip

\noindent
$Organization \ of \ the \ Paper$

The organization of this paper is as follows:
In section 2 we review the Hitchin construction
for Calabi-Yau threefolds and formulate
the problem of  describing
the  moduli space of Riemann surfaces in
terms of the cohomology classes of 1-forms, in the
spirit of Hitchin.
In section 3 we construct a two-dimensional analog
of the  Hitchin functional and
comment on  quantization
of the corresponding theory.
In section 4 we show that this functional can be related to the
gauged  WZW model with a target space an abelian variety.
In section 5 we describe corresponding  quantum theory
and interpret its partition function (which
as a  generating function
for the number of conformal blocks
in $c=2g$ RCFTs) as an entropy functional
on the moduli space of  complex structures.
The non-perturbative coupling
to  two-dimensional  gravity
generates an effective potential on the moduli
space, critical points of which are attractive
and correspond to Jacobian varieties admitting complex
multiplication.
We end in section 6 with conclusions and
discussion on the possible directions for future research.

\newsec{The Hitchin Construction}

The problem of characterizing a
complex manifold in terms
of the  data associated with closed 1-forms on it
goes back to Calabi \Calabi.
In the context of Riemann surfaces, the
non-trivial information encoded
in a closed 1-forms reveals itself in the
ergodicity and integrability of the associated
Hamiltonian systems, which
have been extensively studied  since 1980s by the Novikov school
(see, $e.g.$  \Novikov\ and  the references therein).
Kontsevich and Zorich observed an interesting relation between these
systems and c=1 topological strings~\KZ.
A new twist to the story became possible after Hitchin \refs{\Hgeom, \Hstable}
discovered some diffemorphism-invariant functionals
on   stable $p$-forms.

The critical points of
these topological functionals yield special geometric structures. For example,
the Hitchin functional on 3-forms  defines a complex structure and
holomorphic 3-form in 6 dimensions, and $G_2$ holonomy metrics in
7 dimensions. Hitchin's construction provides an explicit
realization of the idea that
geometrical structures on a manifold can be
described via the cohomology class of a closed form on this manifold.
In this approach, geometric structures arise as solutions
to the equations obtained by extremizing canonical topological action.

In this section we review   Hitchin's approach
to parameterizing  complex structures on a Calabi-Yau
threefold, and formulate the
problem of describing
the moduli space of genus $g$ Riemann surfaces in a similar manner.

\subsec{Stable forms in Six Dimensions }

Let us describe the Hitchin construction, using a
 Calabi-Yau threefold $M$ as an example.
We present below a Polyakov-like version of the Hitchin
functional \Nikita\ (see also \refs{\TM, \GS}),
although originally it was written in a Nambu-Goto-like form
\refs{\Hgeom, \Hstable}. The reason why we need the
Polyakov-like version is
that it is quadratic in fields, and
therefore is more suitable for quantization.
We will also extend the construction of \Nikita\ in order to
incorporate the generalized geometric structures \HitGen.

Let us introduce  a (stable) closed poliform $\rho$, which is
a formal sum of the odd differential forms
$\rho= \rho^{(1)}+ \rho^{(3)}+ \rho^{(5)}$
on a compact oriented six-dimensional manifold $M$.
If we fix the cohomology class $[\rho]$ of this poliform,
it  defines a generalized Calabi-Yau  structure on $M$
as follows. Consider the functional
\eqn\hgcy{
S=- {\pi \over 2} \int_{M} \Big(
\sigma(\rho) \wedge \CJ^{\varsigma\upsilon} \Gamma_{\varsigma\upsilon}  \rho +
\sqrt{-1} \lambda \, \tr (\CJ^2 + \Id)
\Big),
}
where  $\sigma \big(\rho^{(k)} \big) = (-)^{[k/2]} \rho^{(k)}$
transforms the standard wedge pairing between the differential forms
into the Mukai pairing \Gualtieri,
the $6$-form  $\lambda$ serves as a Lagrange multiplier, ~~~~~~~~~~~~~
$\varsigma, \upsilon = 1, \ldots, 12$ are indices in $TM \oplus T^* M$,
the matrix
$\Gamma_{\varsigma\upsilon}  = [\Gamma_{\varsigma}, \Gamma_{\upsilon} ]$ is
defined by  the gamma-matrices $\Gamma_{\varsigma}$ of $\rm Clifford(6,6)$,
and the tensor field $\CJ \in \End(TM \oplus T^* M)$.
After solving  the equations of motion and
using the constraint imposed by $\la$, this field
becomes a  generalized almost complex structure on $M$:
$\CJ^2 = -\Id $. Hitchin \HitGen\ proved that
this almost  complex structure is integrable
and can be used
to reduce the structure group of $TM \oplus T^* M$ to
$SU(3,3)$. This endows $M$ with a generalized Calabi-Yau
structure.

It is perhaps more illuminating to see how this construction
gives rise to the ordinary Calabi-Yau structure,
when $\rho$ is a stable closed 3-form: $\rho = \rho^{(3)}$.
The Polyakov-like version  \refs{ \Nikita} of the  Hitchin functional
has the form\foot{The coefficient $-\pi / 2$ in front of the
integral can  be
fixed after comparing the Hitchin action with the
black hole  entropy functional  (see, $e.g.$, \refs{\TM, \GSV}).
It is tempting to speculate that this
normalization factor can also be determined from the topological
considerations, similarly to
the way the coefficient $-{1/8 \pi}$ in front of the
WZW functional is fixed.}
\eqn\hcy{
S=- {\pi \over 2} \int_{M} \Big(
\rho \wedge \imath_{\CK} \rho + \sqrt{-1} \lambda \, \tr (\CK^2 + \Id)
\Big).
}
Here   $\CK \in \End{T_{\IR} M}$ is a
traceless vector valued 1-form. We denote it as
$\CK$ in order to distinguish it from
the generalized complex structure $\CJ$.
Also, $\rho=\rho^{(3)}$ is a closed 3-form
in a fixed de Rham cohomology class.
It can be decomposed as
$\rho = [\rho] + d \beta$,  where $[\rho] \in H^{3}(M, \IR)$ and
$\beta \in \Lambda^{2} T^* M$.
The equations of motion, obtained by varying  $\b$ in  the
functional \hcy, accompanied by the closeness condition for $\rho$,
take the form
\eqn\rclsd{
d \rho = 0, \qquad d \, \imath_{\CK} \rho =0.
}
The Lagrange multiplier $\lambda$ imposes
the constraint $\tr \CK^2(\rho) = -6$
on the solution of the equation of  motion for the field $\CK$,
which in some local coordinates on $M$ can be written as
$\CK_{\ a}^{b} \sim \epsilon^{b a_1 a_2 a_3 a_4 a_5 }
\rho_{a_1 a_2 a_3} \rho_{a_4 a_5 a} $.
This allows to identify $\CK(\rho)$ as an almost complex structure:
$\CK^2(\rho) = -\Id$.
Moreover, it can be shown \Hgeom, that this almost complex structure
is integrable.
Therefore, solutions of \rclsd, parameterized by
the cohomology class $[\rho]$, define
a unique holomorphic  3-form on the Calabi-Yau manifold $M$,
according to
\eqn\omhol{
\Omega = \rho + \sqrt{-1} \imath_{\CK(\rho)} \rho.
}
We can use the periods of \omhol\
to introduce local  coordinates on the complex
moduli space of Calabi-Yau. Then, a holomorphic 3-form $\Omega$,
viewed as a function
of the cohomology class $[\rho]$, gives a map between
an open set in $H^{3}(M, \IR)$ and
a local Calabi-Yau moduli space \Hgeom.
We will call it the Hitchin map.
After integrating out the field $\CK$ we arrive at the original
Hitchin functional \Hgeom, written in the Nambu-Goto-like form:
\eqn\hro{
S=- {\pi \over 2} \int_{M} \rho \wedge *_{\rho} \, \rho.
}
Here $*_{\rho}$ denotes the Hodge star-operator for the
 Ricci-flat K\"{a}hler metric on $M$, compatible with the
 complex structure $\CK(\rho)$.
Finally, the value of the Hitchin functional \hcy\ calculated
at the critical point \rclsd\ can also be written
in terms of the holomorphic 3-form \omhol\  as follows:
\eqn\homega{
S=-i { \pi \over 4 } \int_{M} \Omega \wedge \bar \Omega.
}
%

\subsec{ Riemann Surfaces and Cohomologies of 1-forms}

We want to find an analog of the Hitchin construction for
the two-dimensional surfaces.
It is natural to expect that the role  that was played by
the closed 3-forms in $3_{\IC}$ dimensions, in  $1_{\IC}$ dimensions
will be played by the  closed 1-forms.
Therefore, we want to construct a functional,
depending on closed 1-forms in a fixed cohomology class,
critical points of which
will determine the complex structure on
 a genus $g$ Riemann surface  $\Si$.
In fact, a two-dimensional  version of the Hitchin functional was already
discussed in  \refs{\Nikita, \GS}.
There, it was pointed out that it
is very similar to  Polyakov's formulation of the
bosonic string as
a sigma model coupled to the two-dimensional
gravity.

However, before discussing the explicit form of
this functional, we want to explain  why a naive
carry-over of the Hitchin idea from six to two dimensions
will not work. First, the very existence
of the Hitchin map is  based on the fact that
in the case of Calabi-Yau threefold $M$ the dimension of the
intermediate cohomology space $\dim H^{3}(M,\IR)$
coincide with the dimension of the
moduli space of calibrated  Calabi-Yau manifolds\foot{The
calibrated Calabi-Yau manifold is a pair: $(M, \Omega)$, where
$M$ Calabi-Yau threefold and  $\Omega$ is a fixed non-vanishing
holomorphic 3-form on $M$. Hitchin construction naturally gives
calibrated Calabi-Yau manifolds.},
which is  equal to $ 2 + 2 h^{2,1}$.
In the case of a genus $g$ Riemann surface
$\dim H^{1}(\Sigma_g, \IR) = 2g$,
but dimension of the moduli space $\CM_g$ for $g>1$
is $\dim\CM_g=6g-6$.
Therefore, the cohomology class of a closed 1-form on $\Si$
does not contain enough data to describe the moduli space.
This is,  of course, not surprising,
as it is well known
that natural parameterization of the moduli space $\CM_g$
is given in terms of the Beltrami differentials $\mu$,
which are dual to the holomorphic quadratic differentials
$\chi \in H^{0}(\Si, \Omega^{\otimes 2})$. In particular,
$\dim H^{0}(\Si, \Omega^{\otimes 2})= 6g-6$, as it should be.
One could then try to use $H^{0}(\Si, \Omega^{\otimes 2})$
 instead of  $H^{1}(\Sigma_g, \IR)$, but
if we go by this route, we will lose the "background independence"
on the complex structure on $\Si$,
which is a nice feature of the Hitchin construction.
The relevant set-up in this case seems to be
provided by the theory of beta-gamma systems  \refs{\Nikitac,\Nikitabg}.
However, it turns out that it is hard to write down an analog
of the Hitchin functional  for  $(\mu, \chi)$ system
with decoupled conformal factor.

The only exception, when the dimension of the moduli space
coincides with the dimension of the first cohomology space, is
the  elliptic curve $\Sigma_1$,
which is in fact  a direct one-dimensional analog of the Calabi-Yau
threefold. In this case, $\dim H^{1}(\Sigma_1, \IR) = 2=\dim\CM_1$
and therefore we might expect that one closed 1-form
can play the role of $\rho$ in two dimensions.
However, as it was noted in \Nikita, one needs at least two
closed 1-forms in order to write down a two-dimensional analog of the
Hitchin functional.
In a certain sense, it is a lower dimensional artefact, as there
just don't happen to be enough indices to write down a non-zero expression.

Clearly, some modification of the Hitchin construction
is needed in the two-dimensional case.
We suggest the following extension that
preserves the
spirit of the original construction.
First, we will use  complex cohomologies instead of the real ones:
\eqn\frst{
H^1(\Si, \IR) \to H^1(\Si, \IC).
}
Second, for a genus $g$ surface we will consider $g$ closed 1-forms:
\eqn\sec{
H^1(\Si, \IC) \to \big(H^1(\Si, \IC)\big)^{\otimes g}.
}
As we will see, this  will allow us  to
define close analog of the Hitchin functional.

\newsec{ Construction of the Lagrangian }

The case of interest for us in this section is a complex valued
1-forms on a two dimensional compact surface $\Si$ of genus $g$.
We will not assume that $\Si$ is  endowed with any additional
structures, such as a metric or  complex structure.
Instead, in the spirit of Hitchin, we would
like to construct a functional, critical points of which will
define a complex structure on $\Si$, making it a Riemann
surface.
In order to keep the presentation self-contained and to fix the notations,
we start with a brief review of the basics of Riemann surfaces.
Then we proceed to the construction
of the functional on the space of closed 1-forms, the
critical points of which in a fixed cohomology class
are harmonic 1-forms. The complex structure on $\Sigma_g$
will arise from the cohomology classes
of these 1-forms. We will also briefly discuss the quantization
of the corresponding theory.

\subsec{Mathematical Background on Riemann Surfaces }

We summarize below some basic facts from the theory of compact
Riemann surfaces \refs{\ACGH,\FK}. Let $\Si$ be a topological
surface with $g$ handles, that is a compact connected oriented
differentiable manifold of real dimension 2. The number of handles
$g$ is the genus of $\Si$. Topologically, $\Si$ is completely
specified by the Euler number $\chi(\Si) = 2-2g$. In particular,
the dimensions of the homology groups are
\eqn\scoh{
\dim H_{0}(\Si)=1, \quad \dim H_{1}(\Si)=2g, \quad \dim H_{2}(\Si)=1.
}
On $\Si$ one can choose the canonical symplectic basis of  1-cycles
$\{A_I, B_I\}, \ I=1,\ldots,g$
for $H_{1}(\Si)$, with the intersection numbers
\eqn\abint{
\#(A_I, A_J) = 0, \quad \#(A_I, B_J) = \delta_{IJ}, \quad \#(B_I, B_J) = 0.
}
This basis, however, is not unique. The ambiguity is controlled by the
Siegel modular group $\G_g = \Sp(2g,\Z)$ preserving symplectic pairing \abint:
\eqn\abmod{\eqalign{
\pmatrix{ B \cr A}\ \to \ \pmatrix{ B \cr A}' =
\pmatrix{a & b \cr c & d} \pmatrix{ B \cr A},
\qquad \pmatrix{ a & b \cr c & d} \in \Gamma_g
}}
Once we have chosen a homology basis  $\{ A_I, B_I\}$, or ``marking" for $\Si$,
we can cut surface along $2g$ curves homologous to the canonical basis and get a
$4g$-sided polygon with appropriate boundary identifications.
This representation of $\Si$ in terms of the polygon is
very helpful in deriving some important identities.
For example, one can show that for any closed
1-forms $\eta$ and $ \theta$ on $\Si$
\eqn\rid{
\int_{\Si} \eta \wedge \theta =  \sum_{I=1}^{g}
\Big( \oint_{A_I}\eta \oint_{B_I} \theta - \oint_{A_I} \theta
\oint_{B_I} \eta \Big),
}
which is called the Riemann bilinear identity.
The scalar product of two closed
1-forms $\eta$ and $ \theta$ on $\Si$ is given by the
Petersson inner product:
\eqn\scp{
\langle \eta , \theta \rangle  = {i \over 2}\int_{\Si} \eta \wedge \bar \theta.
}
As follows from the Riemann bilinear identity,
this scalar product depends only on the cohomology class
of the closed forms: $\langle \eta , \theta \rangle =\langle [\eta] , [\theta] \rangle $.
The canonical  symplectic form
\eqn\symp{
\CS( \eta , \theta)  = \int_{\Si} \eta \wedge \theta
}
for  closed 1-forms  also depends only on the cohomology class:
$\CS( \eta , \theta)  =\CS ([\eta] ,[\theta])$.

Let us introduce the basis  $\{ \a^I,\b^I \}, \ I=1,\ldots,g$ for
$H^1(\Si, \IR)$, which is dual to the canonical homology basis \abint:
\eqn\abbas{\eqalign{
\oint_{A_I} \a^J = \d^{IJ}, \qquad \oint_{B_I} \a^J = 0 \cr
 \oint_{B_I} \b^J = \d^{IJ}, \qquad \oint_{A_I} \b^J = 0.
}}
The ambiguity of this basis is controlled by an exact 1-forms on $\Si$.
Therefore, we can think of $\{ \a^I,\b^I \}$ as of some fixed
representatives in  the de Rham cohomology class. A natural way
to fix this ambiguity is to  pick some Riemann metric $h$ on $\Si$
and require $\{ \a^I,\b^I \}$ to be harmonic:
\eqn\abharm{
d *_{h} \a^I = 0, \quad d *_{h} \b^I = 0
}
where $*_{h}$ is a Hodge $*$-operator defined by $h$.
This choice provides  a canonical basis for $H^1(\Si, \IR)$,
associated with the metric $h$.
We will always use Euclidean
signature on $\Si$.

Topological surface $\Si$ endowed with a complex structure
is called a Riemann surface. Let
us recall that an almost complex structure on $\Si$ is
a section ${\rm J}$ of a vector bundle
$\End (T_{\IR} \Si)$ such that ${\rm J}^2 = -1$.
Here $T_{\IR} \Si$ is a real tangent bundle of
$\Si$. If we pick some (real) local coordinates
$\{x^{a}\},\ a=1,2$ on $\Si$, then ${\rm J}$ can
be represented by a real tensor field which
components ${\rm J}^{a}_{\ b}$ obey
\eqn\Jcomp{ {\rm J}^{a}_{\ b}{\rm J}^{b}_{\ c} = - \d^a_{\ c} }
Here and in what follows, a sum over
the repeating indices is always assumed.
We reserve the indices $\{{a,b,c,\ldots}\}$ that range from $1$ to $2$,
for the world-sheet (Riemann surface),
 and  indices $\{{I,J,K,\ldots}\}$  that range from $1$ to $g$,
for the complex coordinates on the target  (first cohomology) space.
The indices $\{{i,j,k,\ldots}\}$
label real coordinates on  the target space and range from $1$ to $ 2g$.
We do not distinguish between the upper and lower indices. In particular,
we do not use any metric to contract it.
We will also sometimes omit indices and use matrix notations
in the  target space
for shortness.

According to the Newlander-Nirenberg theorem ${\rm J}$, is an
integrable complex structure if it is covariantly constant:
\eqn\Jcons{ \nabla_a{\rm J}^{b}_{\ c} =0. }
In fact, any almost complex structure
on a topological surface is integrable, and therefore below we will just call
it  a complex structure.
In particular, we will be interested in a complex structure compatible with the
metric $h$. In  local coordinates the metric has the form
\eqn\gmet{
h= h_{ab} dx^a \otimes dx^b,
}
and the corresponding complex structure is given by
\eqn\Jg{
{\rm J}(h)^a_{\ b} = \sqrt{\det\| h_{df}\| } \,  \e_{bc}  h^{ca},
}
where $\e_{11}=\e_{22}=0$, $ \e_{12}=-\e_{21}=1$. It is straightforward
to check that this complex structure indeed obeys \Jcomp-\Jcons.
Notice that complex structure \Jg\  depends only on the conformal
class of the metric, since it is invariant under the conformal transformations:
\eqn\Jconf{
h \to e^{\varphi}h, \quad {\rm J}(h) \to {\rm J}(h).
}
The complex coordinates ${z, \bar z}$ on $\Si$ associated with \Jg\ are
determined from the solution of the Beltrami equation
\eqn\belt{
{\rm J}^a_{\ b} {\p z \over \p x^a} = i {\p z \over \p x^b}.
}

Given a marking for $\Si$, there is a unique
basis of holomorphic abelian differentials of the first
kind $\om^I \in H^0(\Si, \Omega)$,
normalized as follows
\eqn\abpers{
\oint_{A_I} \om^J  = \d^{IJ}.
}
Here $\om^I=\om^I_z dz$. Holomorphic 1-differentials
span $-i$ eigenspace of the Hodge $*$-operator
for the metric compatible with the complex structure:
\eqn\hsconv{\eqalign{
* \om = -i \om, \cr
* \bar \om = +i \bar \om.
}}
The period matrix of $\Si$ is defined by
\eqn\prmat{
\tau^{IJ} = \oint_{B_I} \om^J.
}
If we apply the Riemann identity \rid\
to the trivial 2-form $\om^I \wedge  \om^J = 0$, we find that
the period matrix is symmetric:
\eqn\rrels{
\tau^{IJ} =\tau^{JI}.
}
The imaginary part of the period matrix can be represented as follows:
\eqn\imt{
\i \, \tau^{IJ} = {i\over 2} \int_{\Si} \om^I \wedge \bar \om^J.
}
If we use the fact that the norm \scp\ of the non-zero holomorphic
differentials  of the form  $\nu = \nu_I \om^I$
is  positive: $\langle \nu, \nu\rangle >0$, we find
that the period matrix has a positive definite imaginary part:
\eqn\rrelim{
\i \, \tau >0.
}
Now we can express the holomorphic abelian differentials \abpers-\prmat\
via the canonical cohomology basis \abbas\ of harmonic 1-forms \abharm\
on $\Si$ as follows
\eqn\hadf{
\om = \a + \tau \b,
}
where we used the matrix notations.
Under the modular transformations \abmod\ the period matrix
transforms as
\eqn\tmod{
\tau \ \to \ \tau' = (a\tau + b)(c\tau + d)^{-1},
}
while the basis of abelian   differentials  transforms as
\eqn\omod{
\om \ \to \ \om' =  \big(\tau c^T +d^T \big)^{-1} \om.
}

The space of a complex $g \times g$ matrices obeying \rrels, \rrelim\
is the Siegel upper half-space $\CH_g$.
We will call it the Siegel space, for short.
Torelli's theorem states that
a complex structure of $\Si$ is uniquely defined by the period matrix
up to a diffeomorphism. Moreover, to each complex structure there
corresponds a unique point in the fundamental domain
of the modular group
\eqn\fdm{
\CA_g =  \CH_g / \G_g.
}
Unfortunately,  for  higher
genus surfaces the converse is not true (Schottky's problem).
This is easy to see, since for $g>3$ the dimension of \fdm\
 $\dim_{\IC} \CA_g = {g(g+1) \over 2}$
is bigger than
the dimension of the complex structures moduli space   $\dim_{\IC}{\CM}_g =3g-3$.

\subsec{The Canonical Metric }

There is a canonical K{\"{a}}hler metric on a  Riemann surface, the so-called
Bergmann metric. Sometimes  it is also called the Arakelov metric in  literature.
It can be written in terms of the
abelian differentials \abpers\ as
\eqn\bmet{
h_{z\bar z}^{\rm B}= \big( \i \, \tau \big)_{IJ}^{-1} \, \om^I_z \bar \om^{ J}_{\bar{z}}.
}
This metric has a nonpositive curvature. If $g \geq 2$, the curvature
vanishes at most in a finite number of points, and by an
appropriate conformal transformation \bmet\ can be brought into a metric
of constant negative curvature (see, $e.g.$ \Jost).
The K{\"{a}}hler form  corresponding  to the Bergmann metric is given by
\eqn\bkal{
\varpi^{\rm B}={i\over 2}\big( \i \, \tau \big)_{IJ}^{-1} \, \om^I \wedge   \bar \om^J.
}
It is easy to see that the volume of the Riemann surface
in this metric is independent of the complex structure
and is equal to the genus:
\eqn\knorm{
\int_{\Si} \varpi^{\rm B}=g.
}
The special role of the Bergmann metric will become clear  if we consider
the period map $z \to \xi^I$ from the Riemann surface $\Si$ into its
Jacobian variety ${\rm Jac} (\Si) = \IC^n / (\IZ^n \oplus \tau \IZ^n)$:
\eqn\pmap{
\xi^I = \int_{z_0}^z \om^I.
}
Here $z_0$ is some fixed point on  $\Si$, the exact choice of which is usually
not important. Jacobian variety, being a flat complex torus, is endowed with a
canonical metric, which is induced from the Euclidian metric on $\IC^n$.
The Bergmann metric \bmet\ is nothing but a pull-back of this canonical metric
from ${\rm Jac} (\Si)$ to $\Si$ under the period map
\pmap.

The metric \bmet\  does not depend
on the choice of a basis\foot{In the orthonormal basis
$\{ \omega_o^I = \omega_{oz}^I dz :\
\langle \omega_o^I, \omega_o^J\rangle = \d^{IJ}\}$,
the metric takes the canonical form:
$h_{z\bar z} = \sum\limits_{I=1}^{g} \big|\omega_{oz}^I \big|^2$.}
in a space of  holomorpic
differentials $H^0(\Si, \Omega)$.
In particular, it is invariant  under the modular transformations \tmod-\omod.
If we consider $\om^I$ as a set of $g$ closed 1-forms
on $\Si$ in a fixed cohomology class,
parameterized by the period matrix $\tau \in \CA_g$,  then \bmet\ combined
with \Jg\   gives an explicit realization of the
Torelli's theorem, by providing the map $\CA_g \to \CM_g$.
This should be viewed as a two-dimensional
analog of the Hitchin map  \refs{\Hgeom, \Hstable} from
the  cohomology space  of the stable forms on a compact six-dimensional
manifolds to the moduli space of the calibrated Calabi-Yau threefolds.

Indeed, let us recall that
the space of all metrics on a genus $g$
surface $\Si$ is factorized as follows
\eqn\fact{
{\rm Met}(\Si) =\CM_g \times \Diff (\Si) \times {\rm Conf} (\Si).
}
Once we fixed the cohomology  class of $\om^I$,
we are not allowed to do the conformal transformations,
since this will spoil the closeness of $\om^I$.
Therefore,  expression \bmet\ provides a unique representative
among the conformal structures on $\Si$.
This takes care of the ${\rm Conf} (\Si)$ factor in \fact.
Moreover, since diffeomorphisms do not change the cohomology class,
the Bergmann metric \bmet\ is invariant
under the action of  the $\Diff (\Si)$  group,
and we end up  on the moduli space
of genus $g$ Riemann surfaces $\CM_g$.

\subsec{An analog of the Hitchin Functional in Two Dimensions }

As we discussed earlier, the problem in
defining an analog of the Hitchin functional
in two dimensions is
that the cohomology class of only one 1-form is not
enough to parameterize the moduli space of
complex structures. However, if we take
$g$ closed 1-forms on a genus $g$ surface,
this can be done.
In fact, this will give us even
more degrees of freedom
than we need  ($g^2$ complex parameters instead of  $3g-3$),
but it is a minimal set of data
that we can start with, because of the Schottky problem.
The functional that we will use is
a direct generalization of \refs{ \Nikita, \GS}.
The fields of the theory are
    \item{$\bullet$}  $ \z^{I}:$ \ $g$
    closed complex valued 1-forms, $d \z^{I} =0$
    \item{ $\bullet$} $\CK \, :$ \ real traceless vector valued 1-form,
    $\CK\in \End(T_{\IR}\Si)$
    \item{$\bullet$} $\la \ :$ \ imaginary 2-form

\noindent
The Lagrangian has the form
\eqn\Ht{
L =  {k \pi\over 4} \langle \zeta^I ,  \zeta^{J} \rangle^{-1}
\int_{\Si}  \big( \zeta^I \wedge \imath_{\CK} \, \bar \zeta {^J} +
\bar \zeta{^J} \wedge \imath_{\CK} \,  \zeta {^I}\big)
- i  { k \pi\over 4} \int_{\Si}  \lambda \,  \tr \big( \CK^2 + \Id \big),
}
where $k$ is a coupling constant,
2-form $\la$ serves as a  Lagrange multiplier,  and $\Id$
is a unit $2 \times 2$ matrix. Hermitian $g \times g$ matrix
$\langle \zeta^I ,  \zeta^{J} \rangle^{-1}$ is an inverse of
the scalar product \scp\ for 1-forms.
We assume that   cohomology classes
of 1-forms $[\zeta^I]$ are linear independent.
In order to discuss classical equations of motion for the action \Ht\
and their solutions it is useful to
write is explicitly in components:
\eqn\Htloc{
L = {k\pi\over 4}\int_{\Si}
\Big(\langle \zeta^I ,  \zeta^{J} \rangle^{-1}
\big(\zeta^I_a \bar \zeta{_c^J} + \zeta^I_c \bar \zeta{_a^J} \big)
\CK^c_{\ b} -{i\over 2} \la(x) \e_{ab} \big(\CK_{\ c}^d \CK_{\ d}^c +
2 \big) \Big) dx^a \wedge dx^b,
}
where  $\z^I= \z^I_a dx^a, \quad \CK = \CK^a_{\ b} {\p \over \p x^a}
\otimes dx^b, \quad
\la = \half \la(x)\, \e_{ab} dx^a \wedge dx^b$, and
$\e_{11}=\e_{22}=0$, $ \e_{12}=-\e_{21}=1$.
The equations of motion for $\CK$ give
\eqn\Jsol{
\CK_{\ b}^a = {i \over 2 \la(x)} \langle \zeta^I ,  \zeta^{J} \rangle^{-1}
\big(\zeta^I_b \bar \zeta{_c^J} +
\zeta^I_c \bar \zeta{_b^J}  \big) \e^{ca},
}
where $\e^{11}=\e^{22}=0$, $ \e^{12}=-\e^{21}=-1$, such that
$\e^{ac}\e_{cb} = \d^a_{\ b}$.
Variation with respect to $\la$ imposes the constraint
\eqn\Jcnstr{
\CK_{\ b}^a \CK_{\ a}^b = -2,
}
which is solved by setting\foot{Notice that it is
possible for $\lambda$ to vanish at some points, if the
determinant of the  induced metric becomes zero.
In this case, expression \Jsol\ is not well defined,
and potentially is singular. Later we will see, that
in order to give well-defined complex structure,
the cohomologies $[\z^I]$ should lie in the Jacobian locus
in the Siegel upper half-space.}
\eqn\lsol{
\la(x) = \pm {i} \sqrt{\det\| h_{ab}\| },
}
where, by definition,  $\det\|h_{ab}\| =
\half h_{ab} h_{cd} \e^{ac} \e^{bd}$, and
$ h_{ab}$ is the metric induced on $\Si$:
\eqn\hjmet{
h_{ab} = {1 \over 2} \langle \zeta^I ,  \zeta^{J} \rangle^{-1}
\big(\zeta^I_a \bar \zeta{_b^J} +
\zeta^I_b \bar \zeta{_a^J}  \big).
}
We choose the  ``$+$" sign in \lsol\ by requiring positivity of the
Lagrangian \Htloc\ after solving for $\CK$ and $\la$:
\eqn\lextr{
L = {k\pi \over 2}  \int_{\Si}  \sqrt{\det\| h_{ab}\| }.
}
The corresponding solution
\eqn\Jnxsol{
\CK_{\ b}^a  = { 1
\over \sqrt{\det\|h_{df}\| }} \, h_{bc} \e^{ca} =
- \sqrt{\det\|h_{df}\| }  \, \e_{bc}  h^{ca}
}
represents the action of the complex structure, compatible
with the metric \hjmet, on the cotangent bundle $T^*_{\IR} \Si$:
\eqn\KJ{
\CK = {\rm J}^{-1} = -{\rm J}.
}
This should be compared with \Jg.
Let us introduce the notation $*_{\z}$ for a Hodge star operator,
defined by the metric \hjmet. For example, the Hodge dual of
a 1-form $\th = \th_a dx^a$ is given by
\eqn\stardef{
*_{\z} \th = \th_a \sqrt{\det\|h_{df}\| } \, h^{ab} \e_{bc} dx^c.
}
This  Hodge $*$-operator acts on 1-forms exactly  as the
 field $\CK$ \Jsol:
\eqn\staract{
*_{\z} \th= \imath_{\CK} \th.
}
Therefore,  we can rewrite the action \Ht\ in yet another form:
\eqn\Htor{
L = {k\pi\over  4} \langle \zeta^I ,  \zeta^{J} \rangle^{-1}
\int_{\Si}
\zeta^I \wedge *_{\z} \, \bar \zeta{^J} \ + \ {\rm h.c.}
}
This expression is similar to the
Hitchin functional \hro\ and relation between \Ht\ and \Htor\
is very much like relation between the Polyakov and Nambu-Goto actions
in string theory, as was noted in \refs{\Nikita, \GS}.

Following
the idea of Hitchin, we should restrict this
functional to the closed forms on $\Si$ in a given de Rham cohomology
class, and look for the critical points.
In order to parameterize variations of
$\z^I$ in a fixed cohomology class $[\z^I] \in H^1(\Si, \IC)$, we
decompose it as
\eqn\zfrm{
\z^I = [\z^I] + d \xi^I,
}
where $\xi^I$ is a proper function $\Si \to \IC^g$.
By  varying $\xi^I$ in \Ht, we get
\eqn\zhclsd{
 d *_{\z} \z^I =0.
}
Thus, the critical points of the functional
\Htor\ correspond to the harmonic forms on $\Si$.
The complex dimension of the space
of  harmonic 1-forms on $\Si$ is equal to $g$.
Since initial conditions \zfrm\ are
parameterized by  $g$ linear independent vectors
$[\z^I]$,  solution to \zhclsd\
will give us a basis in
the space of  harmonic 1-forms.
We can  parameterize cohomology classes
$[\z^I]$ using their periods over the $A$
and $B$-cycles:
\eqn\zabpar{
[\z^I] = A^{IJ} \a^J + B^{IJ} \b^J,
}
where  $A^{IJ}$ and  $B^{IJ}$ are $g\times g$ complex
matrices. We  impose
some natural restrictions on the form of these matrices.
First, since the action \Ht\ is invariant under the
linear  transformations
\eqn\glg{
\z^I \to M^{IJ} \z^J,   \quad M^{IJ} \in {\rm GL}(g, \IC),
}
we can always set $A^{IJ} = \d^{IJ}$ by using this
transformation with $M= A^{-1}$.
Then, \zabpar\ becomes
\eqn\zppar{
[\z^I] = \a^I + \Pi^{IJ} \b^J,
}
where $\Pi = A^{-1} B$.  The fact that all
cohomology classes
$[\z^I]$ are linear independent means that
\eqn\rankpi{
\rank \, \Pi =g.
}
The second restriction comes from the fact that
the matrix of the scalar products of 1-forms
\eqn\zmet{
\langle \z^I, \z^J  \rangle= {i \over 2} \big(\Pi^{\dagger}-\Pi \big)^{IJ}
}
should be invertible and positive definite for the theory, based on
the action \Ht, to be well-defined. Moreover,
it is natural to require that cohomology classes  $[\z^I]$
do not intersect\foot{
This condition can be imposed, for example, by adding
a term of the form $i {\rm A}_{IJ} \int_{\Si}\z^I \wedge \z^J $
to the action \Ht\ and integrating out antisymmetric
matrix  ${\rm A}_{IJ}$. This term is purely topological (it is
not coupled to $\CK$ and depends only on the
cohomology classes), so it does not affect
the ordinary Hitchin story.}
\eqn\conintr{
\int_{\Si}  \z^I \wedge \z^J=\Pi^{IJ}-\Pi^{JI}=0 .
}
Let us recall that this intersection number is essentially
the canonical symplectic form $\CS(\z^I , \z^J)$   on $H^1 (\Si, \IC)$,
defined in \symp. Therefore, from the perspective
of future quantization of the cohomology space,
it is necessary to require that the points $[\z^I]$
in the configuration space commute.
This requirement is similar to considering
only commuting set of the periods in quantum mechanics of the self-dual form
(see, $e.g.$, \WitEff).

Therefore, instead of dealing with
all non-degenerate matrices $\Pi \in {\rm GL}(g, \IC)$,
we can concentrate  only on the matrices that obey
\eqn\piresrt{
\Pi^T = \Pi, \quad \i \Pi >0.
}
In other words,  we  parameterize
 cohomology classes $[\z^I]$ by the points
on the  Siegel upper half-space  $\CH_g$.
In fact, $\CH_g$ is
the smallest linear space
where we can embed Jacobian variety $\Jac(\Si)$ without knowing its
detailed description, which is unavailable for $g>4$ because of the Schottky problem.
There is a natural action
of the symplectic group on $\CH_g$.
We will denote this "target" modular group as $\Sp(2g, \IZ)_{\rm t}$,
in order to distinguish it from the "world-sheet" modular
group $\Sp(2g, \IZ)_{\rm ws}$ acting on the cover of the
moduli space of Riemann surfaces.

 Given the solution to \zhclsd,
the complex  structure on $\Si$ is uniquely determined by
the corresponding cohomology class via \hjmet-\Jsol,
very much in the spirit of Hitchin.
The map $ \z^I \to *_{\z} \z^I$ globally
defines a decomposition of $\z^I$ on components of type $(1,0)$
and $(0,1)$ with respect to this complex structure.
For example, the $(1,0)$ component of the solution to \zhclsd
\eqn\abel{\eqalign{
 \z^I_{(1,0)}  = \z^I+ i *_{\z} \z^I,
}}
being a harmonic, must be a linear combination of the abelian
differentials \abpers.
This observation allows us to express the period matrix
as a function of the cohomology classes:
\eqn\taucoh{
\tau^{IJ} = \sum_{K}\Big( \oint_{A_K} \z^I_{(1,0)} \Big)^{-1}
\oint_{B_J} \z^K_{(1,0)}.
}
In practice, however, we will have to solve the
equation  \zhclsd\ in order to compute
corresponding  period matrix via \taucoh.
This should be as hard to do as
to solve the Schottky problem.
Furthermore, the complex structure
\Jsol, that we will get, will in general
 be different from the background complex structure
on the abelian variety $\CT(\Pi)$,
that we use to parameterize the cohomologies.
Only if we start from a point\foot{To be precise,
we can also use any point that can be obtained from this one by the
action of the modular group $\Sp(2g,\IZ)_{\rm t}$.}
on the Siegel space that corresponds to the Jacobian variety
$\CT(\Pi) =\Jac\big(\Si(\tau)\big)$, the critical
point of the functional \Ht\ will give us the
same complex structure on the world-sheet
as on the target space.
In this case harmonic maps \zhclsd\ are
promoted to the holomorphic maps, and  \taucoh\ gives $\tau =\Pi$.
The metric  \hjmet\ then is the Bergmann metric \bmet.
This will happen on a very rare occasion, since Jacobian locus  has measure zero
in the Siegel space.
However, in general there is no obstruction
for the map $\CH_g \to \CM_g$ defined by \taucoh,
since all almost complex structures on $\Si$ are integrable.

Formally, this is the end of the ordinary Hitchin
story in two dimensions. However, a new interesting direction
for study
emerges if we allow the cohomology classes $[\z^I]$ to vary.
In this case   we will be dealing with the effective
quantum mechanics of $g$ points  on the Siegel space $\CH_g$\
defined by the functional \Ht.

Let us discuss the dependence of this  functional  on
the "massless" degrees of freedom encoded in $\Pi$ and $\CK$.
We choose some  complex structure
on $\Si$, which is equivalent to fixing the corresponding value
of the field $\CK$.
Modulo diffeomorphisms,
it is defined by the corresponding period matrix $\tau$.
This is equivalent\foot{We assume that some marking for $\Si$ is fixed,
and discuss  the modular group $\Sp(2g,\IZ)_{\rm ws}$ issues later.} to
choosing a set of the
abelian differentials of the first kind \abpers.
Then, we can express the
cohomology class \zppar\ as follows:
\eqn\zpom{
[\z] = (\Pi - \bar \tau) {1\over \tau - \bar \tau} \om -
(\Pi -  \tau) {1\over \tau - \bar \tau} \bar \om.
}
The background dependence
on the complex structure on $\Si$ is encoded in the
period matrix~$\tau$.
Let $*$ be the Hodge star-operator compatible
with this complex structure: $\imath_{\CK} \to *$.
Using the identity
\eqn\instfor{
\z^I \wedge * \, \bar \z{^J} = i \z^I \wedge \bar \z{^J}
- {i \over 2} \big(\z^I - i * \z^I \big) \wedge
{\big( \bar \z{^J} + i *  \bar \z{^J} \big)}
}
and assuming that the classical equations of motion
\zhclsd\ are satisfied, we get the following expression
for the functional \Ht:
\eqn\sptplus{\eqalign{
L(\Pi, \tau) =  {kg\pi} + {k\pi\over 4}
\Tr {1\over \i \Pi} (\Pi - \tau)  {1\over \i \tau}
\bar{(\Pi -\tau)}.
}}
It is clear that this expression has a maximum at
the point  $\Pi  = \tau$  on the Siegel plane.
Moreover, it is
straightforward to check,
using the symmetry of the matrices $\Pi$ and $\tau$,
and the positivity of $\i \tau$, that
\eqn\pitau{
\Pi  = \tau
}
is the only solution\foot{There is also a
nonphysical solution $\Pi = \bar \tau$, that does not
lie on  the Siegel upper-space, since $\i \Pi<0$ in this case.}
of the corresponding equation of motion
\eqn\lpiteq{
 {\p L(\Pi, \tau)\over \p \Pi}= 0.
}
Therefore,  if we allow the cohomology classes  in
the theory with Lagrangian \Ht\ to
fluctuate, we find the following picture.
For the generic period matrix $\Pi$, parameterizing
the cohomology classes, solution
to the equation of motion \zhclsd\ for the "massive"
degrees of freedom (scalars $\xi^I$ in \zfrm) give harmonic maps
$\z: \Si \to \CH_g$.
Further extremization
with respect to $\Pi$ picks up only the
holomorphic maps that correspond
to the Jacobian variety  $\Jac\big(\Si(\tau) \big) \in \CH_g$
of the Riemann surface $\Si(\tau)$ with the
period matrix $\tau$.

Another important feature of  the expression
\sptplus\ is that it is  invariant under the
diagonal subgroup of the group
$\Sp(2g,\IZ)_{\rm t}\times  \Sp(2g,\IZ)_{\rm ws} $,
which acts as follows
\eqn\tpimod{\eqalign{
\Pi  \to  \Pi'   = & (a\Pi + b)(c\Pi + d)^{-1} \cr
\tau \ \to  \tau' = &(a\tau + b)(c\tau + d)^{-1}.
}}
This can be easily checked using the basic relations
for the $\Sp(2g,\IZ)$ matrix $\pmatrix{ a & b \cr c & d}$:
\eqn\spprop{\eqalign{
ad^T-bc^T = a^T d-c^T b = & \one_{g\times g} \cr
ab^T-ba^T=cd^T-dc^T=& 0 \cr
a^Tc-c^Ta=b^Td-d^Tb=& 0.
}}
This, in particular, implies that after
integrating $e^{-L(\Pi,\tau)}$ over the
Siegel space $\CH_g$ we will get a modular invariant function
of $\tau$.
As we will see shortly, this gives an interesting
topological quantum mechanical toy model on $\CM_g$.
However, this toy model  can hardly be interpreted
from the entropic principle perspective.
Therefore, further refinement of the functional \Ht\
will be needed.

\subsec{Towards the Quantum Theory}

Consider the following partition function
defined by the functional \Ht:
\eqn\pfnve{
Z_g(k,\tau) = \int {\CD \Pi  \CD \xi \over \det \i \Pi}\,  e^{-L},
}
where the canonical  modular invariant measure is  used.
It is assumed that we have fixed the value of the field $\CK$,
corresponding to
the complex structure on a Riemann surface $\Si(\tau)$
with the period matrix $\tau$.
After performing the
Gaussian integral over $\xi$ and  using \sptplus, we find
\eqn\pfnveex{
Z_g(k,\tau) =  e^{-{k g\pi}}\int  \CD \Pi
\,  \exp \Big( {-{k\pi\over 4}
\Tr {1\over \i \Pi} (\Pi - \tau)  {1\over \i \tau}
\bar{(\Pi -\tau)}} \Big).
}
The canonical modular invariant measure on $\CH_g$ is
\eqn\hgminv{
\CD \Pi = \big(\det \i \Pi \big)^{-g-1}
\prod_{I\leq J}^g \big| d \Pi_{IJ}\big|^2
}
Since we know that the exponent in \pfnveex\
 has only one minimum  \pitau, for large $k$
we can study the perturbative expansion of the matrix integral
\pfnveex\ near $\Pi = \tau$. It is convenient to
describe the fluctuations by introducing the matrix $H$ as
follows
\eqn\pih{
\Pi = \tau  - \i \tau H.
}
Then \pfnveex\ becomes
\eqn\perth{
Z_g(k,\tau) =e^{-{kg\pi }} \int  \CD H
\,  \exp \Big( -{k\pi\over 4}
\Tr\, { H \bar{H} {1\over 1 - \i H}} \Big).
}
As we discussed earlier, because of the
invariance of the exponent \sptplus\ in
\pfnveex\ under the diagonal modular action \tpimod,
the partition function
$Z_g(k,\tau)$ is a modular invariant  function of $\tau$.
Therefore, it descends to a function
on the  moduli space $\CM_g$ of genus $g$ Riemann surfaces.
Notice that  expression \perth\ does not depend on $\tau$ at all.
Therefore, the modular invariant function that we will get is
actually a constant\foot{
Here we ignored possible contribution
 from the boundary terms.
On the boundary of the moduli space, when $\det \i \Pi=0 $
and $\det \i \tau = 0 $, the integral \pfnveex\ needs to be
carefully regularised.
This could result
in a non-trivial $\tau$-dependence,
but such effects are  beyond the scope of this paper.
}.

The integral in \perth, as a function of $k$,
can be expressed in terms of
the 1-matrix model.
Let us split the matrix $H$ into its real and imaginary parts
\eqn\himr{
H = H_1 + i H_2.
}
Then
we can rewrite \perth\ as
\eqn\gkgpert{
Z_g(k,\tau) =  e^{-{kg \pi}}\int  \CD H_2
\int  \CD H_1 \,
\exp\Big( -{k\pi\over 4} \Tr \, (H_1^2 + H_2^2) {1\over 1 -  H_2} \Big),
}
where we integrate over the
symmetric matrices, and
the matrix measures are defined in accordance with \hgminv
\eqn\hmeas{
\CD H_1 = \prod_{I\leq J}^g  d H^{IJ}_1, \qquad
\CD H_2 = \big(\det H_2 \big)^{-g-1}
\prod_{I\leq J}^g  d H^{IJ}_2
}
The integral over $H_1$ is Gaussian, and gives
$ ( {2 \over k\pi})^{{g(g+1)\over2}}\det (1 -  H_2)^{{g+1\over2}} $,
up to a numerical constant.
Then, \gkgpert\ becomes
\eqn\gkgpertn{
Z_g(k,\tau) =  \big( {2 \over k\pi}\big)^{{g(g+1)\over2}} e^{-{kg \pi}}\int
\prod_{I\leq J}^g  d H^{IJ}_2
\Big(\det {1 -  H_2\over H_2^2}\Big)^{{g+1\over2}}
e^{ -{k\pi\over 4} \Tr    {H_2^2\over 1 -  H_2} }.
}

It is unclear, however, whether
this expression has any interesting interpretation.
In principle,
we could  use an alternative  definition
of the partition function,
where   the canonical measure is multiplied
by some  function of $\Pi$, instead of \pfnve.
Morally,  this is equivalent to
adding corresponding topological terms
(depending only on the cohomology classes $[\z]$,
without coupling to $\CK$) to the action~\Ht.
This will bring us into the realm of the matrix models.
However, it looks like just a digression to $0_{\IC}+1$
theory, and we are looking for the links
with the higher dimensional theories.

Thus,
we suggest a natural generalization of the theory, that comes
from the following observation.
If we concentrate  only
on the massive modes in the functional \Ht,
described by the free non-compact fields $\xi$,
it looks very much like the action for the
string propagating on a complex torus
$\CT^g_{\IC} = \IC^g/\IZ^g \oplus \Pi \IZ^g$
with the period matrix $\Pi$. Indeed, as we
discussed earlier, the metric $\big(\i \Pi \big)^{-1}$
is a canonical metric induced  on the torus
from the flat Euclidian
metric on $\IC^g$.
Therefore, from the stringy point of view
the non-compact scalars $\xi: \Si \to  \IC^g $ can be promoted
to the  maps $\phi: \Si \to  \CT^g_{\IC}$ with non-trivial winding numbers.
Then, the cohomology class $[\z]$ in the combination
$\z=[\z]+ d\xi$ can be interpreted as a  background abelian
gauge field on the torus: $[\z] \to \CA $.
Therefore, we want to substitute
\eqn\genprop{
\z \to d \phi +\CA
}
in the functional  \Ht\ and study resulting quantum theory.
This modification will give us a new
insight on  the
gauged WZW model for abelian varieties,
coupled to the complex structure
on $\Si$ in the specific way  \Ht.

\newsec{Gauged WZW Model for Abelian Varieties
and the Hitchin Functional}

In this section we argue that in order to use
quantum  theory based on the Hitchin functional for computing
topological invariants, one has to incorporate stringy effects
into it. In particular, the target space
has to be compactified, and in accordance with that
one has to consider topologically non-trivial  maps
$\Si \to \CT^g_{\IC}$,
instead of $\Si \to \IC^g$. Here  $\CT^g_{\IC}$
is a complex $g$-torus, viewed as
a principally polarized abelian variety.
Moreover, the translation group of the target space
has to be gauged, so that a two-dimensional
stringy version of the  Hitchin functional
becomes the  gauged
WZW model with an abelian gauge group.
It is well known  that the
partition function of this model,
representing
the number of conformal blocks in corresponding toroidal
CFT, is independent of the complex structure on $\Si$.
However, as we will see,
in the Hitchin extension of the
model the coupling to the two-dimensional gravity
appears non-perturbatively via the instanton effects.

Before describing the topological extension of  Hitchin functional
in two dimensions,
we will recall some general aspects of the  gauged
Wess-Zumino-(Novikov)-Witten model.
This theory was extensively studied in the literature,
see $ e.g.$
\refs{\GK, \Gaw, \Sglas,  \WitHol, \BT, \Ger, \WitGrass,
\BTKS},
therefore below we just summarize basic features of the model, following
 \refs{\WitHol, \WitGrass,\BTKS}.
We will also use some facts about the
abelian Chern-Simons theories with the gauge
group $U(1)^d$, which  has been discussed
recently   in great details in \refs{\GMMS,\BM}.
We will be particularly interested in the case  $d=2g$,
and focus on viewing gauge group
as a complex algebraic variety.

\subsec{Review of the Gauged WZW Model}

Let $G$ be a compact Lie group. The group  $G$ acts on itself
by left and right multiplication, which is convenient to
view as the action of  $G_L\times G_R$.
For any  subgroup $H_L\times H_R \subseteq G_L\times G_R$
consider a   principal $H_L\times H_R$ bundle $X$
over Riemann surface $\Si$, with connection $(\CA_L, \CA_R)$.
Let ${\rm J}$ be a complex structure on $\Si$.
As we discussed earlier,  it determines the action
of the Hodge $*$-operator, corresponding to the Riemann metric
compatible with this complex structure, on 1-forms.
Consider the functional:
\eqn\gwzw{\eqalign{
I(\CA_L, \CA_R;g) =& -{1\over 8 \pi} \int\limits_{\Si}
\Tr \big(g^{-1}d_{\CA} g \wedge *  \, g^{-1}d_{\CA} g \big) - i \G(g)+\cr
&+ {i\over 4 \pi} \int\limits_{\Si} \Tr \big( \CA_L \wedge dg g^{-1}+
\CA_R \wedge  g^{-1} dg  + \CA_R  \wedge  g^{-1} \CA_L   g \big),
}}
where $d_\CA$ is the gauge-covariant
extension of the exterior derivative:
\eqn\agge{
d_{\CA} g =dg + \CA_L g-g\CA_R,
}
and $\G(g)$ is the topological WZNW term:
\eqn\tptrm{
\G(g)   ={1\over 12 \pi}
\int\limits_{B: \, \p B = \Si} \Tr (g^{-1}dg)^{\wedge \, 3}.
}
Here $\Tr$ is an invariant quadratic form on the Lie algebra $\Lie G$ of
the group $G$,
normalized so that $\G(g)$ is well-defined modulo $2\pi\IZ$.
The field\foot{not to be confused
with the genus $g$ of $\Si$.} $g$ is promoted from the
map $g: \Si \to G$ to a section of the bundle $X\times_{H_L\times H_R} G$,
where $G$ is understood as a trivial principal $G$ bundle over $\Si$.
We are interested in the non-anomalous gauging,
which is only possible if for all $t,t' \in \Lie (H_L\times H_R)$
\eqn\angng{
\Tr_L tt'-\Tr_R tt'=0
}
where $\Tr_L$ and $\Tr_R$ are traces on  $\Lie H_L$ and $\Lie H_R$.
The standard choice for a non-abelian group is $H_L= G_L$ and $H_R = G_R$,
with  diagonal action $g \to h^{-1} g h$. This
gives the $G/G$ gauged WZW model.

Consider the following  propagator
\eqn\lrprop{
\big\langle \Psi_{\CA_L}(\Si)\big|\Psi_{\CA_R}(\Si)\big\rangle =
\int \CD g e^{-k I(\CA_L, \CA_R;g)}.
}
This should be compared to \prop.
In order to simplify the
notations, we will often write $\Psi_{\CA}$ instead
of $\Psi_{\CA}(\Si)$, when
the  dependence on the complex structure of $\Si$ is not
essential.  We will use the notation
$\Psi_{\CA}\big(\Si(\tau)\big)$
if we want to stress    dependence on the complex structure,
parameterized by the period matrix $\tau$ of $\Si$.

By performing the change of   variables $g \to g^{-1}$ in the
functional integral \lrprop,  we find that the propagator has the
necessary property
\eqn\lrprpr{
\bar{ \big\langle \Psi_{\CA_L}\big|\Psi_{\CA_R}\big\rangle} =
\big\langle \Psi_{\CA_R}\big|\Psi_{\CA_L}\big\rangle.
}
Furthermore, using  the Gaussian integration and
the Polyakov-Wiegmann formula
\eqn\pw{
I(0,0,gh) = I(0,0,g) + I(0,0,h) -{1\over 4\pi} \int_{\Si}
\Tr \, g^{-1} dg \wedge dh h^{-1},
}
it is easy to check that propagator \lrprop\  satisfies
the "gluing" condition \pprop.
It is also straightforward to obtain the  relation
\eqn\coc{
I(\CA_L^h, \CA_R^{\tilde h}; h^{-1}g\tilde h)=
I(\CA_L, \CA_R; g) - i \Phi(\CA_L; h)+ i \Phi(\CA_R; \tilde h),
}
where the gauge transformed connection is
\eqn\alrgge{
\CA_L^h =h^{-1}\CA_L h + h^{-1}d h , \qquad
\CA_R^{\tilde h}= \tilde h^{-1}\CA_R \tilde h  + \tilde h^{-1}d \tilde h,
}
and the  cocycles
\eqn\cocdef{
\Phi(\CA; h) = {1\over 4\pi}  \int_{\Si} \Tr \CA \wedge  dh h^{-1} - \G(h)
}
are independent of the
complex structure (metric) on $\Si$, and  satisfy
\eqn\cocrel{
\Phi(\CA; hh')= \Phi(\CA^h; h') + \Phi(\CA; h).
}
Infinitesimal form of these global gauge transformations,
combined with a direct variation over $\CA_{Lz}$ in
the functional integral \lrprop,
leads to the following set of the equations for the propagator
\eqn\pshol{\eqalign{
{D\over D \CA_{Lz}} \,
\big\langle \Psi_{\CA_L}\big|\Psi_{\CA_R}\big\rangle =0
\cr
\Big( D_{a}^L {D\over D \CA_{L a}} +{ik\over 4\pi} \e^{ab} \CF_{ab}^L \Big)\,
\big\langle \Psi_{\CA_L}\big|\Psi_{\CA_R}\big\rangle =0
}}
where we introduced a connection ${D\over D \CA_{L }}$
on the line bundle $\CL^k$ over the space of $H_L$-valued connections
\eqn\dadef{\eqalign{
{D\over D \CA_{L z}}={\d \over \d \CA_{L z}} +{k\over 4\pi} \CA_{L\bar z} \cr
{D\over D \CA_{L\bar z}}={\d \over \d \CA_{L\bar z}} -{k\over 4\pi} \CA_{L\bar z},
}}
and  covariant derivatives on the principal $H_L$ bundle
over $\Si$
\eqn\dordef{\eqalign{
D_{a}^L = \partial_a + [\CA_L,\,.\, ],
}}
with the curvature form
\eqn\fcurv{
\CF^L = [D^L, D^L]= d\CA_L + \CA_L \wedge \CA_L.
}
Connections \dadef\ obey
canonical commutation relation
\eqn\alcomm{
\Big[{D\over D \CA_{L z}(z)}, {D\over D \CA_{L \bar w}(w)}\Big] =
+{k\over 2\pi} \delta(z,w)
}
The propagator \lrprop\ also satisfies a
set of  conjugate equations
that describe its dependence on $\CA_R$.
These equations are obtained from \pshol-\alcomm\
by a change of the indices and signs,
according to
\eqn\ischange{
L \, \leftrightarrow \, R, \qquad +k \, \leftrightarrow \, - k.
}
Geometrically, it means that the propagator
$\big\langle \Psi_{\CA_L}\big|\Psi_{\CA_R}\big\rangle $
is an (equivariant) holomorphic section
of the line bundle $\CLL = \CL^k \otimes \CL^{-k} $.

The quantum field theory with Lagrangian $L= k I(\CA, \CA;g)$, $k \in \IZ_+$
is conformal and gauge invariant and is called
the  $G/G$ gauged WZW model  for the non-abelian group $G$ at level $k$:
\eqn\zggi{
Z_k({G/G};\Si) =
\int {\CD g \CD \CA \over \vol(\rm Gauge)}
e^{- kI(\CA,\CA;g)}.
}
It is a two-dimensional sigma model with target space $G$
gauged by a non-anomalous subgroup ${\rm diag}(G_R\times G_R)$.
The partition function of the $G/G$ gauged
WZW model can be also written as
\eqn\zgg{\eqalign{
Z_k({G/G};\Si) = \Tr_{\CA} \big\langle \Psi_{\CA}(\Si)\big|\Psi_{\CA}(\Si)\big\rangle
=\int {\CD \CA_L \CD \CA_R \over \vol^2(\rm Gauge)} \,
 \Big|\big\langle \Psi_{\CA_L}\big|\Psi_{\CA_R}\big\rangle \Big|^2,
}}
where we used \lrprpr. This should be compared to \trc, with
the identification $Z_k({G/G};\Si) = Z_{\Si \times \S^1}$.
After performing  the Gaussian integration,
applying  the Polyakov-Wiegmann formula
and relation \coc, we indeed get \zggi.

The  gauged WZW functional \gwzw\ allows one to connect
three-dimensional Chern-Simons theory and
its dual two-dimensional rational conformal field theory
in a simple and effective way.
The partition function of the WZW model is
\eqn\pwzw{
Z_k(G;\Si)  =  \int
{\CD g  } e^{-kI(0,0;g)}.
}
The holomorphic factorization
of the WZW model into the conformal
blocks can be explained by  observing \WitHol\ that \pwzw\
can also be written as
\eqn\pwzwh{
Z_k(G;\Si)  =
\langle \Psi_{0}(\Si) | \Psi_{0}(\Si) \rangle=
\big | |\, \Psi_{0}(\Si)\rangle \big|^2=
\int {\CD \CA \over \vol (\rm Gauge)} \,
 \big|\big\langle \Psi_{0}\big|\Psi_{\CA} \big\rangle \big|^2.
}
The WZW model is a
rational conformal field theory if it is constructed from a
finite number of  conformal
blocks\foot{There are many definitions of RCFT,
but this one is the most convenient for our purposes.}.
In this case the
conformal blocks of the  WZW model are in one-to-one correspondence with
the states in a Hilbert space\foot{For simplicity we are not considering marked
points on $\Si$.}, obtained from canonical quantization of the Chern-Simons
theory on $\Si  \times \IR$ \refs{\WitJon, \EMSS}.
A geometrical interpretation of this  Hilbert space (achieved in the framework of the
geometrical quantization \APW) is that it is a space $V_{g,k}(G)$
of (equivariant) holomorphic sections of $k$-th power of the determinant
line bundle $\CL$ over the moduli space of (semistable) holomorphic
$G_{\IC}$-connections  on $\Si$,  which by the  Narasimhan-Seshadri
theorem is the same as
the moduli space $\CM_G$ of flat connections on the
principal $G$-bundle over  $\Si$.
Thus, the Hilbert space is given exactly by the holomorphic sections that satisfy \pshol.

Let $s_{\g}(\CA;\tau)$, $\g=1, \ldots \dim H^0(\CM_G, \CL^{ k})$ be
an orthonormal basis in the space $H^0(\CM_G, \CL^{ k})$ of holomorphic sections.
Then we can write the propagator in \pwzwh\ as
\eqn\psiabas{
\big\langle \Psi_{0}\big( \Si(\tau)\big)\big|
\Psi_{\CA}\big( \Si(\tau)\big) \big\rangle =
\sum\limits_{\g=1}^{\dim H^0(\CM_G, \CL^{ k})}
 \bar{F_{\g} (\tau)} \, s_{\g} (\CA;\tau).
}
The coefficients $F_{\g} (\tau)$ in \psiabas\ are the conformal
blocks of the  WZW model.
Of course, the dimensions of the space $V_{g,k}(G)$ of conformal blocks and
the space $H^0(\CM_G, \CL^{ k})$ of holomorphic sections  coincide.
After plugging \psiabas\ into \pwzwh\
and using the orthonormality of the basis $s_{\g}(\CA;\tau)$, we obtain
\eqn\wzwbas{
Z_k\big(  G;\Si(\tau)\big) = \sum\limits_{\g=1}^{\dim V_{g,k}}
\big|F_{\g}(\tau)\big|^2.
}
The propagator \lrprop\  can now be written as
\eqn\psiaa{
\big\langle \Psi_{\CA_L}\big( \Si(\tau)\big)\big|
\Psi_{\CA_R}\big( \Si(\tau)\big) \big\rangle =
\sum\limits_{\g=1}^{\dim V_{g,k}}
 s_{\g} (\CA_L;\tau)  \bar{s_{\g} (\CA_R;\tau)},
}
which is a unique solution to the equations \pshol\
(and conjugate equations \ischange) obeying
the "gluing" condition \pprop. After
plugging this into \zgg\ we find
\eqn\dimvkg{
 Z_k({G/G};\Si)=  \dim V_{g,k}(G)= \dim H^0(\CM_G, \CL^{ k}).
}
Therefore,
the partition function of the
gauged WZW model computes the
dimension of the  Chern-Simons Hilbert space $V_{g,k}(G)$,
which coincides with the number of conformal
blocks in the corresponding RCFT. This can be viewed as
an example of the universal index theorem \inddef\
for the  universal partition function \trc\
of $\Si\times \S^1$, which in this case is
equal to $Z_k({G/G};\Si)$.
The higher cohomology groups  vanish since we are
dealing with the
integrable representations of RCFT.

Another way to explain \dimvkg\ is to observe
\refs{\Ger,\WitGrass} that
the propagator \lrprop\ is exactly the free propagator
of the Chern-Simons theory multiplied by the projector on
the gauge invariant subspace, enforcing the Gauss law.
In other words, equation \zggi\ is equivalent to
$ Z_k({G/G};\Si) = \Tr \ 1,$ which yields \dimvkg.

From the CFT algebra viewpoint, the number of conformal
blocks $\dim V_{g,k}(G)$ is given by the E.~Verlinde's formula \Verfor.
For example, when $G=SU(2)$,
\eqn\sutver{
\dim V_{g,k}({ SU}(2))= \Big({k+2 \over 2}\Big)^{g-1}
\sum_{j=0}^{k} \sin^{2-2g} {(j+1)\over k+2}\pi.
}
The  gauged WZW model  provides a constructive method of computing the
dimension of the Verlinde algebra
via the localization of the functional integral \refs{\BT, \Ger}.

\subsec{Abelian Case}

We are particularly interested
in the case when $G$ is an abelian group: $G\sim U(1)^{2g}$.
Moreover, we want to view it as an algebraic complex variety
with a fixed complex structure. Therefore, we will
describe  $G$ as  a $g$-dimensional complex torus
$\CT$, which is a principally polarized abelian variety:
$\CT = \IC^g/ \Lambda$, $\Lambda = \IZ^g\oplus \Pi \IZ^g$.
Sometimes we will use  the notation $\CT(\Pi)$
to show the explicit dependence of $\CT$ on the
defining period matrix $\Pi$.

Let us first describe an analog of the functional \gwzw\
for the  case of  abelian group $\CT \sim U(1)^{2g}$.
In complex coordinates, it has the form\foot{
 We should be careful with the expression \gwzw\
in the case of $U(1)^{2g}$ abelian group. Naively, it looks like we can
 take $2g$ copies of the  WZNW term \tptrm\  for $U(1)$
group, but this expression vanishes for abelian group element $g=e^{i \varphi}$.
The analog of this term  in the
abelian group case is
$
\G(\phi) = { \pi \CG_{IJ}} \int_{\Si} d\phi^I \wedge  d\bar \phi{^J},
$
which can be interpreted  a $B$-field.
It is  crucial for
a global identification with the corresponding three-dimensional
Chern-Simons theory.
}
\eqn\agwzw{\eqalign{
I(\CA_L, \CA_R;\phi) =  {\pi \CG_{IJ}\over 2} \int\limits_{\Si}
 d_{\CA} \phi^I \wedge *  \, d_{\bar \CA} \bar \phi{^J}
& + i\big(\CA_L^I + \CA{_R^I} \big)  \wedge  d \bar \phi{^J} +
i \big(\bar \CA{_L^I} + \bar\CA{_R^I} \big) \wedge  d \phi{^J} \ - \cr
& -  i {\pi \CG_{IJ}\over 2} \int\limits_{\Si}
\big( \CA_L^I  \wedge  \bar \CA{_R^J}+\bar \CA{_L^I}  \wedge  \CA{_R^J}\big)
\ -  \ i \G(\phi),
}}
where $(\CA_L, \CA_R)$ are connections on a principal bundle
$\CT_L \times \CT_R$ over $\Si$,  and the scalar fields
$\phi \sim \phi + \IZ + \Pi \IZ$ describe the maps
$\Si \to \CT$, which, after coupling to the gauge fields, are
promoted to the corresponding sections, with the covariant derivative
defined as
\eqn\abgge{
d_{\CA} \phi^I =d\phi^I + \CA_L^I - \CA{_R^I}.
}
The role of the trace operator $\Tr$ in \gwzw\ now is
played by the matrix $\CG_{IJ} = \big({1\over \i \Pi }\big)_{IJ}$,
that defines a canonical metric on $\CT$:
\eqn\mcant{
\CG(\phi, \phi) ={1\over 2}\CG_{IJ}(\phi^I \otimes \bar\phi{^J}+
\bar \phi{^I} \otimes \phi{^J}).
}
The analog of the topological WZNW term is
\eqn\agwzwan{
\G(\phi) = {\pi \CG_{IJ}} \int_{\Si} d\phi^I \wedge  d\bar \phi{^J},
}
which obey the corresponding
Polyakov-Wiegmann formula
\eqn\pw{
\G(\phi+\psi)= \G(\phi) + \G(\psi) +{\pi \CG_{IJ}}  \int_{\Si}
  d\phi^{I} \wedge d\bar  \psi{^J} - d\bar \phi{^{I}} \wedge d \psi{^J}.
}
Under the small gauge transformations
\eqn\gatrab{
 \qquad \CA_L^{\psi} =\CA_L  + d \psi , \qquad
\CA_R^{\tilde \psi}= \CA_R  + d \tilde \psi,
}
the change of the functional \agwzw\
depends on $\CA_{L,R}$ but not on $\phi$  or  the complex structure  of $\Si$:
\eqn\cocab{\eqalign{
I\big(\CA_L^{\psi}, \CA_R^{\tilde \psi}; \phi+ \tilde \psi -\psi \big) &-
I\big(\CA_L, \CA_R; \phi\big)  = \cr  = &
\, i{\CG_{IJ}\pi\over 2} \int_{\Si}
\CA{_R^I} \wedge d\bar{\tilde\psi}{^J} +\bar\CA{_R^I} \wedge d \tilde \psi^J
- \CA_L^I \wedge d \bar\psi{^J} - \bar \CA{_L^I} \wedge d \psi^J.
}}
This should be compared to \coc.
There are certain restrictions \refs{\EMSS, \BN} on the possible choice
of the period matrix $\Pi$ of the torus $\CT(\Pi)$.
First, in order for the functional \agwzw\ to be a well defined modulo
$2\pi i \IZ$, the lattice $\Lambda = \IZ^g\oplus \Pi \IZ^g$
has to be integral. Second, modular invariance requires $\Lambda$
to be even lattice. Therefore,
\eqn\veidf{
\IZ^g\oplus \Pi \IZ^g \in \G_{2g}^{2\IZ}.
}
where $\G_{2g}^{2\IZ}$ denotes the moduli space of
even integral $2g$-dimensional lattices.
The dual conformal field theory in this
case is rational.

Let us define the propagator as
\eqn\abprop{
\big\langle \Psi_{\CA_L}\big|\Psi_{\CA_R}\big\rangle =
\int \CD \phi \, e^{-k I(\CA_L, \CA_R;\phi)}.
}
Notice that  interactions in  \agwzw\ are such that $\phi$ is coupled
 only to $\CA_{L\bar z}$, via the term $d \phi^I \wedge (i-*) \bar \CA{_L^J}$, and to
$\CA_{Rz}$, via the term $d \phi^I \wedge (i+*)\bar \CA{_R^J}$.
Moreover, "left" and "right" gauge fields interact only via
the coupling $\CA_{Rz}^I \wedge \bar \CA{_{L\bar z}^J}$, and
its complex conjugate. This observation,
combined with the Ward identity, that follows from \cocab,
leads to the following set of
equations\foot{We treat $\CA_{L,R}$
and $\bar \CA{_{L,R}}$ as independent variables, and there is also a
corresponding set of equations
with $\CA_{L,R} \to\bar \CA{_{L,R}}$.}, which the propagator obeys:
\eqn\abpshol{\eqalign{
{D\over D \CA_{Lz}^I} \,
\big\langle \Psi_{\CA_L}\big|\Psi_{\CA_R}\big\rangle =0\cr
{D\over D \CA_{R\bar z}^I} \,
\big\langle \Psi_{\CA_L}\big|\Psi_{\CA_R}\big\rangle =0
\cr
\Big( \p_{a} {D\over D \CA_{L a}^I} +{ik\pi} \CG_{IJ} \e^{ab} \p_a \CA_{Lb}^J \Big)\,
\big\langle \Psi_{\CA_L}\big|\Psi_{\CA_R}\big\rangle =0\cr
\Big(  \p_{a} {D\over D \CA_{R a}^I} -{ik\pi} \CG_{IJ} \e^{ab} \p_a \CA_{Rb}^J \Big)\,
\big\langle \Psi_{\CA_L}\big|\Psi_{\CA_R}\big\rangle =0
}}
Here we introduced a connection
\eqn\abdadef{\eqalign{
{D\over D \CA_{L z}^I}={\d \over \d \CA_{L z}^I} +{k\pi}  \CG_{IJ} \CA_{L\bar z}^J,\qquad
{D\over D \CA_{L\bar z}^I}={\d \over \d \CA_{L\bar z}^I} -{k \pi} \CG_{IJ}\CA_{L\bar z}^J, \cr
{D\over D \CA_{R z}^I}={\d \over \d \CA_{R z}^I} -{k\pi}  \CG_{IJ} \CA_{L\bar z}^J,\qquad
{D\over D \CA_{R\bar z}^I}={\d \over \d \CA_{R\bar z}^I} +{k\pi} \CG_{IJ}\CA_{L\bar z}^J,
}}
on the line bundle $\CL^k\times \CL^{-k}$ over the space $\bf A$ of $\CT_L\times\CT_R$-valued
connections on $\Si$.

The geometrical interpretation  of  the equations \abpshol\
is very simple. We pick a standard complex structure on the space  $\bf A$
of connections induced from the complex structure on $\Si$.
In this complex structure, $\CA_{L\bar z}$ and $\CA_{Rz}$ are holomorphic, and
 $\CA_{Lz}$ and $\CA_{R\bar z}$ are antiholomorphic. Then
 the propagator $\big\langle \Psi_{\CA_L}\big|\Psi_{\CA_R}\big\rangle $
 is a holomorphic section of the line bundle $\CLL = \CL^k \otimes \CL^{-k} $,
 equivariant with respect to the action of the abelian group  $\CT_L\times\CT_R$.

It is well-known (see, $e.g.$, \refs{\APW, \Tyurin}) that the basis in
the corresponding space $H^0(\CM_{\CT},\CL^k)$ of the
gauge invariant  holomorphic sections of
 $\CL^k$ is provided by the level $k$ Narain-Siegel theta-functions
$\Theta_{\g}(\CA;\tau|{\Lambda},{k})$, associated with the lattice
$\Lambda$, that defines the  torus $\CT = \IC^g/ \Lambda$.
We will not need an explicit expression for $\Theta_{\g}(\CA,\tau|{\Lambda},{k})$
(it can be found, for example,  in \refs{\GMMS,\BM}).
What is important for us is that the linear independent
Narain-Siegel theta-functions
are labelled by the index $\g\in (\Lambda^*/k\Lambda)^{\otimes g}$,
where $\Lambda^*$ is the dual lattice.
From the viewpoint of the three-dimensional abelian Chern-Simons
theory, these theta-functions are exactly the wave-functions:
$\Psi_{\g} ( \CA;\tau) \sim\Theta_{\g}(\CA;\tau|{\Lambda},{k})$.
Therefore, the dimension of the corresponding Hilbert space
is
\eqn\dmabhs{
\dim Hilb_{\rm CS}({\Lambda},{k}) =\big|\Lambda^*/k\Lambda\big|^{ g}.
}

We can repeat the steps that we did in the non-abelian case,
and connect abelian Chern-Simons theory and its dual CFT via the
functional \agwzw\ and the propagator \abprop.
The property of the functional \agwzw\
\eqn\abconjpr{
\bar{I(\CA_L, \CA_R;\phi) } =I(\CA_R, \CA_L;-\phi)
}
guarantees that the propagator \abprop\ is hermitian:
\eqn\abprop{
\bar {\big\langle \Psi_{\CA_L}\big|\Psi_{\CA_R}\big\rangle} =
\big\langle \Psi_{\CA_R}\big|\Psi_{\CA_L}\big\rangle,
}
since we can always change the variables $\phi \to -\phi$ in the
functional integral \abprop. Moreover, it is straightforward
to show that the propagator obeys the gluing condition \pprop
\eqn\abglid{
\big\langle \Psi_{\CA_L}\big|\Psi_{\CA_R}\big\rangle =
\int {\CD \CA \over \vol({\rm Gauge})}
\big\langle \Psi_{\CA_L}\big|\Psi_{\CA}\big\rangle
\big\langle \Psi_{\CA}\big|\Psi_{\CA_R}\big\rangle,
}
by performing the Gaussian integral over $\CA$,
using the Polyakov-Wiegmann formula, and the fact that
$\int \CD \phi =  \vol({\rm Gauge})$.
This allows us to write down the following
expression for the propagator in terms of the Narain-Siegel theta-functions:
\eqn\abproprstr{
\big\langle \Psi_{\CA_L}\big|\Psi_{\CA_R}\big\rangle =
\sum_{\g\in (\Lambda^*/k\Lambda)^{\otimes g}}
\Theta_{\g}(\CA_L;\tau|{\Lambda},{k})
\bar{ \Theta_{\g}(\CA_R;\tau|{\Lambda},{k})}
}

The partition function of the gauged
WZW model for  abelian group
$\CT$  at level $k$  is defined as
\eqn\zgg{
Z_k ({\CT/\CT};\Si) =\int {\CD \phi \CD \CA \over \vol(\rm Gauge)} \
e^{- kI(\CA,\CA;\phi)} = \Tr_{\CA} \big\langle \Psi_{\CA}\big|\Psi_{\CA}\big\rangle
}
Using \abproprstr\ and  orthonormality of the
Narain-Siegel theta-functions
\eqn\nstaort{
\int { \CD \CA \over \vol(\rm Gauge)}\
\Theta_{\g}(\CA;\tau|{\Lambda},{k})\,
\bar{ \Theta_{\g'}(\CA;\tau|{\Lambda},{k})} = \delta_{\g\g'},
}
it is easy to see that the partition function \zgg\ indeed
computes the dimension of the Chern-Simons theory Hilbert space
\eqn\zggab{
Z_k ({\CT/\CT};\Si) =\big|\Lambda^*/k\Lambda\big|^{ g}.
}
%

\subsec{Hitchin Extension of the Abelian
GWZW Model}

Now we are ready to discuss the Hitchin extension of the
gauged WZW functional \agwzw\ for abelian group.
We want to introduce non-trivial coupling to the complex structure
on $\Si$ by using the operator  $\imath_{\CK}$ instead of the Hodge $*$-operator,
and adding the term $i  \lambda \,  \tr \big( \CK^2 + \one \big) $
to the action.
This leads to  the following functional
\eqn\hgwzw{\eqalign{
I(\CA_L, \CA_R;\phi|\la, \CK)\ =\ &{\CG_{IJ}\pi\over 4} \int\limits_{\Si}
 d_{\CA} \phi^I \wedge \imath_{\CK} \, d_{\bar \CA} \bar \phi{^J} +
 d_{\bar \CA} \bar \phi{^I} \wedge \imath_{\CK} \, d_{ \CA}  \phi{^J} -
 4\sqrt{-1} d \phi{^I} \wedge d  \bar  \phi{^J} +\cr
&+  {\sqrt{-1}\pi\over 2} \CG_{IJ} \int\limits_{\Si}
 \big(\CA_L^I + \CA{_R^I} \big)  \wedge  d \bar \phi{^J}
+\big(\bar \CA{_L^I} + \bar\CA{_R^I} \big) \wedge  d \phi{^J} -\cr
&-  { \sqrt{-1} \pi\over 2}\CG_{IJ} \int\limits_{\Si}
\big( \CA_L^I  \wedge  \bar \CA{_R^J}+\bar \CA{_L^I}  \wedge  \CA{_R^J} \big)
-  {\sqrt{-1}\pi\over 4} \int\limits_{\Si}
\lambda \,  \tr \big( \CK^2 + \Id \big).
}}
The Hitchin extension of  the propagator \abprop\ formally
is given by
\eqn\abprop{
\big\langle \Psi(\CA_L|\la, \CK)\big|\Psi(\CA_R|\la,\CK)\big\rangle =
\int \CD \phi \, e^{-k I(\CA_L, \CA_R;\phi|\la,\CK)}.
}
However, this expression can be interpreted as
a propagator only for the "on-shell" values of the
field $\CK$, such that
\eqn\konsh{
\CK^2 = -\Id.
}
In this case $\CK$ defines a complex structure on $\Si$, and
we can glue together two propagators defined in the same complex
structure, according to the gluing rule \abglid.
Moreover, if we  define a "partition function" as
\eqn\zaform{
 Z_k(\CT|\la,\CK) = \int {\CD \phi \CD \CA\over\vol({\rm Gauge})}
 \, e^{-k I(\CA, \CA;\phi|\la,\CK)},
}
then  formally we can write
\eqn\pertprop{
\int \CD\la \ Z_k(\CT|\la,\CK) =\big|\Lambda^*/k\Lambda\big|^{ g}\
\delta\big(\tr  \CK^2 + 2\big).
}
The meaning of this expression  is that perturbatively the Hitchin extension
\hgwzw\ is equivalent to the ordinary gauged WZW model \agwzw.
There is no non-trivial $\CK$ dependence in \pertprop, and
after performing the integration over $\CD \CK$ we will
 just get some multiplicative constant, depending on $g$.
 This should not be surprising. After all,
 the gauged WZW model computes the
number of conformal blocks (the dimension of the
corresponding Hilbert space), and this number does not
depend on the choice of the complex structure on $\Si$,
which is controlled by the field $\CK$.

However, the very new feature of the Hitchin extension
is that dependence on  $\CK$ can be
restored  non-perturbatively.
Indeed, if  the action $kI(\CA, \CA;\phi|\la, \CK)$
has non-trivial critical point, we
have to do expansion around this point in the
functional integral.
In this case, the answer will depend on the value $\CK_{min}$ of
the complex structure tensor at the minimal point  of the action.

\subsec{Attractor Points and Complex Multiplication}

Therefore, we have to study the
critical points of the functional
\eqn\abprh{\eqalign{
I(\CA, \CA;\phi|\la, \CK)\ & =\ {\CG_{IJ}\pi\over 4} \int\limits_{\Si}
 \big( d\phi^I \wedge \imath_{\CK} \, d \bar \phi{^J} +
 d \bar \phi{^I} \wedge \imath_{\CK} \, d  \phi{^J} -
 4\sqrt{-1} d \phi{^I} \wedge d  \bar  \phi{^J} \big) +\cr
&+  {\sqrt{-1}\pi} \CG_{IJ} \int\limits_{\Si}
 \big( \CA^I   \wedge  d \bar \phi{^J}
+\bar \CA{^I}  \wedge  d \phi{^J} \big)
-  {\sqrt{-1}\pi\over 4} \int\limits_{\Si}
\lambda \,  \tr \big( \CK^2 + \Id \big).
}}
Let us recall that in the functional integral \zaform\
we integrate over the exact parts $\varphi^I$ of  the fields
 $d\phi^I = [d\phi^I] + d\varphi^I$
and sum over non-trivial maps
 $[d\phi^I] \in H^{1}(\Si, \Lambda)$.
After dividing by the gauge transformations, we need to
integrate only over the space of gauge inequivalent flat
 gauge fields $\CA^I \in  H^{1}(\Si, \IC^g)/H^{1}(\Si, \Lambda)$.
Therefore, in the functional \abprh\  the
exact part of $d\phi$  couples only to the term
$ \imath_{\CK} \, d \bar \phi$. By varying $\bar \varphi$,
we get a classical equation of motion, analogous to  \zhclsd:
\eqn\pexag{
d \imath_{\CK} \, d  \phi{^I} = 0.
}
After solving the constraint $\tr  \CK^2 = -2$,
imposed by the Lagrange multiplier $\la$,
the equations of motion for $\CK$ give
\eqn\Kpsol{
\CK^a_{\ b} = {\CG_{IJ} \over 2 \sqrt{\det \| h\|}} \ \big(
 d\phi^I_b  d \bar \phi{^J_c} +
 d\bar \phi{^I_c}   d \phi{^J_b}
 \big)\e^{ca},
}
where $h$ is the metric induced on $\Si$. If we recall that
  $\CG_{IJ} = \big({1\over \i \Pi }\big)_{IJ}$,
this metric takes the form
\eqn\hpjmet{
h_{ab} =  \Big({1\over 2 \i \Pi }\Big)_{IJ}\big( d\phi^I_a  d \bar \phi{^J_b} +
 d\bar \phi{^I_b}   d \phi{^J_a}
  \big).
}
Expressions \Kpsol\ and \hpjmet\ should be compared with \Jsol\ and \hjmet.

For generic choice of the  matrix $\Pi \in \CH_g$ and cohomology vectors
$[d \phi{^I}]\in H^{1}(\Si, \Lambda)$
expression   \Kpsol\ for the complex structure
 can be singular at some points on $\Si$.
Those are  the points where the determinant of the metric \hpjmet\
vanishes\foot{For example,  the metric $h_{z\bar z} = |\om^1_z|^2$
vanishes at zeroes of the abelian differential $\om^1$. Strictly speaking,
even in this singular case it is possible to define  complex structure
globally on $\Si$ via appropriate conformal transformation and analytical
continuation,  but the resulting complex structure will not
be given by  \Kpsol.}.
However, it is easy to find a family of non-singular solutions  \Kpsol.
Let us compare \hpjmet\ with the
expression  \bmet\ for the canonical Bergmann metric on the
Riemann surface $\Si(\tau)$
\eqn\mbetc{
h^{\rm B}_{ab} =  \Big({1\over 2 \i \tau}\Big)_{IJ}\big( \om^I_a \bar \om{^J_b}+
\bar \om{^I_b}  \om{^J_a} \big).
}
The complex structure on $\Si(\tau)$ is defined by the period matrix  $\tau$,
and is such that the differentials $\om^I$ are holomorphic.
If we set
\eqn\onlysol{
 d\phi^I = \om^I,
}
and choose the torus $\CT$, for which
\eqn\ipiitau{
 \Pi = \tau,
}
then the metric \hpjmet\ coincides with the
Bergmann metric:
 $h =h^{\rm B}$, and therefore the complex structure defined by $\CK$
coincides with the complex structure defined by $\tau$.

In order to parameterize  general   non-singular
complex structure solutions \Kpsol\ we
proceed as follows.
Suppose that some $\CK$, given by
\Kpsol, provides a globally well-defined complex structure on $\Si$.
All complex  structures on $\Si$ are parameterized by the
period matrices. Therefore, there is
the period matrix $\tau$ that defines the same
complex structure on $\Si$ as $\CK$.
Then $\CK$  must be equal to the corresponding
Bergmann complex structure: $\CK = \CK^{\rm B}(\tau)$,
which is a canonical complex structure compatible with the
metric $h^{\rm B}$ \mbetc. This gives
\eqn\kombid{\eqalign{
 \sqrt{\det \| h\|^{\rm B}} \Big({1\over  \i \Pi}\Big)_{IJ}   \big(
 d\phi^I_b  d \bar \phi{^J_c} +
 d\bar \phi{^I_c}   d \phi{^J_b}
 \big) =
 \sqrt{\det \| h\|} \Big({1\over  \i \tau}\Big)_{IJ}
 \big( \om^I_b \bar \om{^J_c}+
\bar \om{^I_c}  \om{^J_b} \big).
}}
Since the 1-forms $d \phi$ are harmonic, we
can express them in terms of the abelian differentials:
\eqn\phimnn{
d \phi = M \om + N \bar \om,
}
where $M$ and $N$ are certain $g \times g$ complex
matrices representing  non-trivial mappings $\Si(\tau) \to \CT(\Pi)$.
If $A$ and $B$
are the period matrices of the 1-forms:
\eqn\abdphipr{
A^{IJ} = \oint_{A_J} d\phi^I, \qquad
B^{IJ} = \oint_{B_J} d\phi^I,
}
then
\eqn\mndfs{
M = (B - A\bar \tau) {1\over \tau - \bar \tau}, \qquad
N = - (B- A  \tau) {1\over \tau - \bar \tau}.
}
We stress that \phimnn\ is an exact expression for the 1-forms $d \phi$,
that solves  classical equations of motion,
as opposed to \zpom, that captures only the cohomology class. Once  the
cohomology class $ [d\phi] $ of the 1-forms is fixed, the exact part
$d\phi - [d\phi]$ is uniquely determined by \pexag, which
states that $d\phi$ is a linear combination of the harmonic representatives.

Combining the $\om_b^I \om_c^J$ terms in \kombid, we find
\eqn\mpinnot{
M^T {1\over  \i \Pi} \bar N = 0,
}
which means that either $N=0$ or $M=0$, since $\i \Pi$ is
non-degenerate.
The terms of the form $\om_b^I\bar \om{_c^J}$ give
\eqn\mnoobr{
\Tr \Big( \om_b^T M^T {1\over  \i \Pi} \bar M \bar \om{_c} \Big)
+ \Tr \Big( \bar \om{_b^T}  N^T {1\over  \i \Pi} \bar N \om{_c} \Big)
 =\sqrt{\det \| h\|\over \det \| h^{\rm B}\|} \
 \Tr \Big( \om_b^T {1\over  \i \tau} \bar \om{_c} \Big).
}
Thus, the only way to satisfy
\mpinnot-\mnoobr\ is to set $N=0$, and
\eqn\ptonmt{
M^T {1\over  \i \Pi} \bar M = {1\over  \i \tau}.
}
This equation implies that $\det M \not  = 0$,
since the matrices $\i \Pi$ and $\i \tau$ are not degenerate.
Moreover, in this case we also have $h = h^{\rm B}$.
From \mndfs\ we see, that the condition $N=0$ is equivalent to
\eqn\beqatau{
B = A\tau,
}
so that
\eqn\meqa{
M = A.
}
The columns of the matrix $A$ \abdphipr\ are the vectors
of the lattice $\Lambda = \IZ^g \oplus \Pi \IZ^g$. We
can write it as
$A = P_{\IZ} +   \Pi Q_{\IZ}$, where  $P_{\IZ}$ and $Q_{\IZ}$ are
integral $g\times g$ matrices.
Therefore, the complex structures $\CK_*$ corresponding to the
critical points of the functional \abprh\ can be parameterized by the
period matrices $\tau$, obeying
\eqn\trsrtr{
{1\over  \i \tau} =
\big(P_{\IZ}^T +  Q_{\IZ}^T \Pi  \big) {1\over  \i \Pi}
\big(P_{\IZ} + \bar \Pi Q_{\IZ} \big).
}
This equation puts additional constraint on the
period matrix, which according to \beqatau\ can
be written as  $\tau = A^{-1}B$.
Since the columns of the matrix $B$ are also the vectors
of the lattice $\Lambda$, we
can write it as
$B = P_{\IZ}' + \Pi Q_{\IZ}'$, where  $P_{\IZ}'$ and $Q_{\IZ}'$ are
integral $g\times g$ matrices. Then \beqatau\ takes the form
\eqn\beqatint{
\tau ={1\over P_{\IZ} + \Pi Q_{\IZ}}
\big(P_{\IZ}' + \Pi Q_{\IZ}'\big).
}
Equations \trsrtr-\beqatint\
can be interpreted as a two-dimensional analog
of the attractor equations \refs{\FKS,\Strominger,\CWM,\Moore}.
In the $3_{\IC}$-dimensional case
attractor equations define complex structure of the Calabi-Yau
threefold in terms of the integral cohomology class,
given by the magnetic and electric charges of the associated black hole.
In $1_{\IC}$-dimensional case equations \trsrtr - \beqatint\
define the complex structure of the Riemann surface $\Si(\tau)$
in terms of the integral  matrices $P_{\IZ}, Q_{\IZ}, P_{\IZ}',$ and $Q_{\IZ}'$.

The critical points \Kpsol\ minimize the value
of the functional \abprh, viewed as a function on the moduli
space of complex structures.
Indeed, the second variation of the functional at the critical point is
\eqn\svf{
{\d^2 I(\CA, \CA;\phi|\la, \CK) \over \d \CK^2}\Big|_* =
-{i\pi\over 2 }\la_* =
{\pi\over 2 } \sqrt {\det \| h^{\rm B}\|} >0.
}
If we perform  functional integration
over $\CD \CK$ with the weight $e^{-k I(\CA, \CA;\phi|\la, \CK)}$,
the main contribution will come from these critical points.
Therefore,  from the point of view of the corresponding quantum mechanical
problem on the moduli space of  complex structures,
these points are attractive. We will denote
a set of these points on the moduli space of genus $g$
Riemann surfaces as ${\rm Attr }_g$.

For the particular choice $P_{\IZ}= Q_{\IZ}' = \one$, and $Q_{\IZ} = P_{\IZ}'=0 $
the attractor equations \trsrtr-\beqatint\
reduce to \onlysol-\ipiitau. This allows us to
generate all solutions to \trsrtr-\beqatint\
from \onlysol-\ipiitau\ by an appropriate
symplectic transformation.
Indeed, a compatibility of \trsrtr\ and \beqatint,
combined with the symmetry requirement $\tau^T = \tau$
imposes certain restrictions on the possible choice of the
integer matrices.
After some algebra  one finds that these restrictions are
equivalent to relations \spprop\ for the symplectic group,
with the identification: $a= Q_{\IZ}'^T, \ b=  P_{\IZ}'^T,\
c=Q_{\IZ}^T,\ d= P_{\IZ}^T$.
Therefore,
\eqn\pqmod{\eqalign{
\qquad \pmatrix{ Q_{\IZ}'^T &  P_{\IZ}'^T \cr
\cr
Q_{\IZ}^T & P_{\IZ}^T} \in \Sp(2g,\IZ),
}}
and all solutions to \trsrtr-\beqatint\ for
a given $\Pi$ correspond to the same Riemann surface,
with a different choices of the symplectic basis.
To summarize, we find that  the critical points of the functional \abprh\
on the moduli space of complex structures $\CM_g$
are given by the intersection of the Jacobian locus $\Jac(\Si) \subset \CH_g$
with a set $\Gamma_{2g}^{2\IZ}$ of  abelian varieties generated by the even integer
$2g$-dimensional lattices:
\eqn\att{
{\rm Attr }_g =\Jac(\Si) \cap   \Gamma_{2g}^{2\IZ}.
}

There is  another interesting property of the critical points
defined by \beqatau\ and \trsrtr: the corresponding Riemann surface
$\Si(\tau)$ admits a non-trivial endomorphism, known as
the complex multiplication (CM). The notion of the
complex multiplication appears in the study of black hole attractors and
rational conformal field theories
(see, $e.g.$, \refs{\Moore, \GV} for
more details and references).
In particular, it was shown in \GSV, that
the critical attractor points of the Calabi-Yau holomorpic volume
functional \homega\
(which is morally the higher-dimensional analog of the functional \abprh)
lead to the abelian varieties (associated
with  the coupling constant matrix)
admitting complex multiplication.

In order to illustrate the CM-property of the critical points \Kpsol,
we will use a simple criterion \GV\
that says that an abelian variety  defined by
the period matrix $\tau$  admits complex multiplication,
if $\tau$ obeys a second order matrix equation
\eqn\tcmdf{
\tau n \tau + \tau m-n'\tau-m'=0,
}
for some integer  $g \times g$ matrices $m,n,m',n'$, with
 $\rank( n) = g$.
It is straightforward to show that any solution
to the attractor equations \trsrtr -\beqatint\
also obeys the CM-type equation \tcmdf.
After using \mndfs, the equation  \ptonmt\ takes the form
\eqn\batpi{
\big(B^T - \bar \tau A^T\big) {1\over \i \Pi}
\big(\bar B- \bar A  \tau\big) =
4 \i \tau.
}
By substituting $\bar \tau = \tau - 2 i \i \tau$ into
the real part of \batpi, we find
\eqn\tcmpr{
\tau \r \big(A^T {1\over \i \Pi} \bar A \big)\tau-
\tau\r \big( A^T {1\over \i \Pi} \bar B\big) -
\r \big(B^T {1\over \i \Pi}  \bar A\big) \tau  +
\r \big(B^T {1\over \i \Pi} \bar B\big)= 0,
}
where we used \beqatau\ and the attractor equation \trsrtr\ in the
form  $ A^T {1\over \i \Pi} \bar A  = {1\over \i \tau}$.
Let us now recall that $\Lambda=\IZ^g \oplus \Pi \IZ^g$
is an even integral lattice.
This  guarantees that the corresponding
three-dimensional abelian Chern-Simons
theory is well-defined, and the
associated two-dimensional conformal field theory is rational.
Therefore, for any two vectors ${\bf a,b} \in \Lambda$
a scalar product, defined as
\eqn\scprdf{
({\bf a,b}) = \r \big({\bf a}^I  ({\i \Pi})_{IJ}^{-1} {\bf \bar b}{^J}\big)
}
is an integer, and the norm of any vector of the lattice
$\Lambda$ is an even number:
\eqn\intg{
({\bf a,b}) \in \IZ, \ {\bf a \not = b}; \qquad  ({\bf a,a}) \in 2\IZ.
}
We can write the period matrices \abdphipr\ in terms of the
lattice vectors, as
$A = ({\bf a}_1, \ldots {\bf a}_g)$, and $B = ({\bf b}_1, \ldots {\bf b}_g)$.
Then all elements of the matrices
\eqn\cmfindfg{\eqalign{
n =\r \big(A^T {1\over \i \Pi} \bar A\big), \, \qquad
m \, = - \r \big(A^T {1\over \i \Pi} \bar B \big), \cr
n' = \r \big(B^T {1\over \i \Pi}  \bar A\big), \qquad
m' = -\r \big(B^T {1\over \i \Pi} \bar B\big),
}}
according to \intg\
are integral, and thus equation \tcmpr\ is indeed
of the CM-type \tcmdf. This  fact should not be surprising.
As was proven recently in \Chen,
the complex multiplication on abelian variety is
equivalent to the existence of the rational K\"{a}hler metric.
This is, of course, true in our case, since
we consider abelian varieties  generated
by the even integer lattices.
The fact that the associated CFT in this case is rational,
 fits nicely with the observations of Gukov and Vafa \GV.

\newsec{Quantization and the Partition Function}

In this section, we define a
generating function for the dimension of the
space of conformal blocks in
a family of toroidal $c=2g$ RCFTs on a genus $g$ Riemann surface.
We  use   Hitchin construction to introduce
 coupling   to two-dimensional
gravity. The universal index theorem
in the context of the Chern-Simons/CFT correspondence
is a computation of the number of
conformal blocks via the  gauged WZW model.
After coupling to  two-dimensional gravity
it gives, according to the entropic principle,
the effective entropy functional on the moduli space of complex structures.
The functional is peaked at the attractor points.
We will be interested in the
fluctuation of the complex geometry
around the gravitational instanton solution
corresponding to these points.
It gives some version of the two-dimensional
Kodaira-Spencer theory of gravity.

\subsec{Generating Function for the Number of Conformal Blocks  and Attractors}

We learned that the Hitchin extension $I(\CA, \CA;\phi|\la, \CK)$  of the abelian
gauged WZW model gives rise to  effective potential
on the moduli space of the complex structures,
whose critical points \att\ correspond to Jacobians
of Riemann surfaces admitting
complex multiplication. In order to describe all
such points we have to sum over all even integer lattices
$\{\Lambda(\Pi) = \IZ^g \oplus \Pi \IZ^g: \ \Lambda(\Pi) \in \G_{2g}^{2\IZ}\}$.
This discrete sum is basically a sum over the moduli space\foot{To be more
precise, the moduli space of a $2g$-dimensional torus is
$SO(2g,2g)\over SO(2g)\times SO(2g)\times SO(2g,2g,\IZ)$.
We are interested in the subspace of the complex algebraic tori $\Sp(2g)\over U(g) \times \Sp(2g, \IZ)$
in this moduli space, and moreover, consider only the tori generated by the even integral lattices.} of the
toroidal rational two-dimensional conformal field
theories:
\eqn\gfnvdf{
Z_{g,k}(\Theta|\la, \CK) = \sum_{\Pi :\, \Lambda(\Pi) \in \G_{2g}^{2\IZ}}
\ e^{i \Tr \Theta \Pi}
\int {\CD \phi \CD \CA \over \vol({\rm Gauge})}
e^{-kI(\CA, \CA;\phi|\la, \CK)}.
}
We perform a sum with the weight factor
$\exp({i \Tr \Theta \Pi})$, where $\Theta$
is an auxiliary symmetric matrix,  so that
$Z_k(\Theta|\la, \CK)$
can be interpreted  as a
generating function  capturing all the relevant information about the theory.
In principle, we can go
one step further and sum over the theories at  different levels $k$ as well:
\eqn\sk{
Z_{g}(q,\Theta|\la, \CK)  =\sum_{k=1}^{\infty} q^k Z_{g,k}(\Theta|\la, \CK).
}
If we compute  \sk\ at any  classical value $ \CK_*: \  \CK_*^2 = - \Id$,
the functional $I(\CA, \CA;\phi|\la, \CK_*)$  describes
the ordinary gauged WZW model, and therefore
\sk\ becomes
a generating function\foot{The simplest example of such generating function
corresponds to the $U(1)_k$ theory, describing
the free boson at $k$ times the self-dual radius.
The holomorphic wave-functions of the
dual Chern-Simons theory are the level $k$ Jacobi
theta-functions. The dimension of the corresponding Hilbert space is $k^g$.
Therefore, on this case
%
$Z_g(q) = \sum_{k=1}^{\infty} q^k k^g = Li_{-g}(q)$.
%
}
for the number of conformal blocks in
$c=2g$ RCFTs
\eqn\gfnvdf{
Z_k(q,\Theta|\la, \CK_*)  = \sum_{k} \sum_{\Pi :\, \Lambda(\Pi) \in \G_{2g}^{2\IZ}}
q^k e^{i \Tr \Theta \Pi}
\big|  \Lambda^*(\Pi)/k \Lambda(\Pi)\big|^{g}.
}

Let us discuss the quantum aspects
of the theory on the moduli space of the complex structures
that arises after averaging  the generating function \gfnvdf\ over
the fluctuations of the fields $\CK$ and $\la$,
according to
\eqn\Zg{
Z_{g,k}(\Theta) = \int {\CD \CK  \CD \la   \over\vol\big(\Diff(\Si)\big)}
Z_{g,k}(\Theta|\la, \CK).
}
The measure for the   vector-valued
1-form $\CK$  can be
 defined as follows.
Let us first notice that for any 1-form $\theta$ on $\Si$
\eqn\thktr{
\theta \wedge \imath_{\CK-\tr\CK} \theta =
\theta \wedge \imath_{\CK} \theta -(\tr\CK )\, \theta \wedge \imath_{\Id} \theta =
\theta \wedge \imath_{\CK} \theta,
}
since $\theta \wedge \imath_{\Id} \theta  = \theta \wedge  \theta =0$.
Therefore, $\tr \CK $
does not couple to the scalars $\phi$ in the action \abprh.
This is the reason why we can integrate only over
the traceless tensor fields $\tr \CK = 0$.
In this case the measure on the space of
the fields $\CK $ is induced from the following
metric:
\eqn\kmetp{
\| \d \CK\|^2 = \int_{\Si}    d^2x \, \big(\tr \CK^2\big)^{-{3\over 2}}\,
\tr \big(\imath_{\CK}  \d \CK\big)^2.
}
In order to motivate this choice of the metric,
we  note that
on-shell, $\CK$  is linearly related to the
Riemann metric $h$ on $\Si$: $\CK^a_{\ b} \sim h_{bc} \e^{ca}$.
Traceless vector-valued
1-form $\CK^a_{\ b}$ contains 3 local degrees of freedom,
the same amount as the symmetric metric tensor $h_{ab}$.
However, $\CK$ and $h$ scale differently
under the conformal transformations.
This can be taken care of  by introducing a
conformal factor $\sigma$, such that
\eqn\kgrel{
\CK^a_{\ b} ={h_{bc} \over \sqrt{\det h } } e^{\sigma} \e^{ca}.
}
Then it is easy to see
that the metric \kmetp\ for the variations of $\CK$
that does not involve change of $\tr \CK^2$,
coincides with the standard metric  \Polyakovb\ on the
space of Riemann  metrics
\eqn\hmetp{
\| \d h\|^2 = \int_{\Si}    d^2x \sqrt {\det h} \,  h^{ac} h^{bd} \d h_{ab} \d h_{cd},
}
for the variations of $h$ that does not involve conformal
transformations. In order to parameterize
general variations, we follow the standard procedure \BK,
and introduce complex coordinates on $\Si$, in terms
of which the metric takes the conformal form
$h = h_{z\bar z} dz \otimes d\bar z$. The group
$\Diff(\Si)$  is generated by the coordinate transformations
$z \to z + \varepsilon (z,\bar z)$.
Then the metric \kmetp\ takes the form
\eqn\kmetinf{
\| \d \CK\|^2 = \int_{\Si} d^2x e^{\sigma}
\big( (\d \sigma)^2 +\p  \bar \varepsilon
\bar  \p  \varepsilon \big) +
\sum_{i,j=1}^{3g-g} \d m_i \big(N_2^{-1} \big)^{ij}\d \bar m_j
}
where $m_i$ are coordinates on the moduli space of Riemann surfaces $\CM_g$,
and $N_2$ is the matrix of scalar products
of the quadratic holomorphic differentials on $\Si$.
Therefore, the measure in the functional integral \Zg\ is
given by
\eqn\gk{
\CD \CK = \vol\big(\Diff(\Si)\big) {\det' \Delta_{-1} \over \det N_2} d\sigma  dm
}
where $\Delta_j$ denotes the Laplace-Beltrami
operator acting on the space of the holomorphic $j$-differentials,
 $N_j$ is the matrix of scalar products
of holomorphic $j$-differentials,
and the volume form on the moduli space
is $dm  = \prod\limits_{i=1}^{3g-g} d m_i \wedge  d \bar m_i$.
We see that $\sigma$ plays the role of
 the Liouville field (the conformal factor of the metric).
In particular, we can compute the $\sigma$-dependence of the
determinant in \gk\ using the standard formula
\eqn\ind{
{\det' \Delta_{j} \over \det N_j  \, \det N_{1-j}} =
\big|\det \bar \p_j \big|^2 \,
e^{-{c_j \over 12 \pi} S_L[\sigma]},
}
where $c_j = 6j^2-6j+1$ and $S_L[\sigma]$ is the Liouville action.
However, because of the choice of the parameterizaton \kgrel, the
conformal field $\sigma$ enters the Hitchin extension of the
gauged WZW model \abprh\ in a special way.
The relevant terms of the functional \abprh\ have the form:
\eqn\sdepf{
{\CG_{IJ} \pi} \int_{\Si} d^2x  e^{\sigma}
\p \phi^I \bar \p \, \bar \phi{^{J}} + {i\pi \over 2}
 \int_{\Si} \la ( e^{\sigma}-1)
}
The additional factor $e^{\sigma}$ makes this
theory at the quantum level very  different  from
   Polyakov's bosonic string. Let us recall that
the quantum theory \Zg\ is defined as an expansion around the
attractor point
\eqn\attexp{
\CK = \CK_* + \d \CK, \quad \la  = \la_* + \d \la.
}
This means that we should expand \sdepf\
around $\sigma=0$.
If we formally do this expansion, in perturbation series
we will encounter  terms of the form
\eqn\badtrms{
\sum_{n>0}  {1\over n!}\int_{\Si} d^2x \sigma^n \big\langle
\p \phi^I \bar \p \, \bar \phi{^{J}} \big\rangle.
}
These terms are singular,
since $\langle \phi(z) \bar \phi(w)\rangle \sim \log |z-w|$, and we
are taking the limit $z\to w, \ \sigma \to 0$.
Therefore, for this theory to make sense, \badtrms\
has to be regularized in some way.

However, in the classical (weak coupling) limit $k\to \infty$
we can ignore this  regularization ambiguity.
If we neglect possible contributions from the boundary
of the moduli space, in this limit
the main contribution
to \Zg\ comes from the attractor points \att:
\eqn\clim{
Z_{g,k}(\Theta)\big|_{k\to \infty} =
 \sum_{\Pi \in {\rm Attr}_g }
 e^{i \Tr \Theta \Pi} \Big(
\big|  \Lambda^*(\Pi)/k \Lambda(\Pi)\big|^{g}
+ \ldots\Big).
}
From the viewpoint of the entropic principle \entrop,
it means that the wave-function \wf\  on the moduli space
of the complex structures $\CM_g$ is peaked
at the attractor points \att.

There is  one physically natural way
to resolve the regularization ambiguity in \sdepf.
We would like to think about the corresponding theory as of a
$1_{\IC}$-dimensional analog of the Kodaira-Spencer theory
of gravity \BCOV.
In the $3_{\IC}$-dimensional case, the
target space KS action \BCOV\ also suffers from the
regularization ambiguities. However, there the topological string $B$-model
provides
a natural regularization.
Unfortunately, the  higher genus  topological string amplitudes vanish
if the target manifold has dimension different from the critical
dimension $\hat c=3$, so we can not view
$1_{\IC}$-dimensional analog of KS theory  as a topological strings on $\Si$.
Instead, we can define it by requiring that a generating function
\Zg\ should be identified with the corresponding computation in the dimensional
reduction of the  Kodaira-Spencer theory
of gravity \BCOV\ from six to two dimensions.

\subsec{Dimensional Reduction of the Topological M-Theory}

Let us  discuss the relation between the
two-dimensional  Hitchin model studied above,
 and  the dimensional reduction
of the topological M-theory\foot{We thank  C.~Vafa and E.~Witten for raising the question
about  the relation between Hitchin functionals in different dimensions.}.
At the moment, there is no consistent quantum definition of
the  topological M-theory \TM. However, many ingredients
of the theory can be identified at the classical level.
In particular,  a seven-dimensional
topological action
\eqn\sdadf{
S_{7} = {1\over 2\pi }\int_{M_7} H \wedge dH,
}
which is a $U(1)$ Chern-Simons theory for 3-form $H$ plays an important role
in interpreting  the topological string partition function as a  wave-function
(see, $e.g.$, \refs{\TM, \Nikita, \GS, \Witbckrnd} and references therein).

On the other hand, it is well-known \Verlinde, that we can get a
$1_{\IC} +1$-dimensional abelian Chern-Simons theory from \sdadf\ via dimensional
reduction on the manifold of the form $M_7=M_4\times\Si \times \IR$.
Using the ansatz
$H = \sum \a_i A^i$, where $\a_i$ are integral harmonic 2-forms on $M_4$, we obtain:
\eqn\dred{
{1\over 2\pi } \int\limits_{M_4\times\Si \times \IR}H \wedge dH
\to {K_{ij}\over 2\pi }\int\limits_{\Si \times \IR}
A^i \wedge d A^j.
}
Here $K_{ij}$ is an
intersection form for harmonic 2-forms
on $M_4$. If we use the spin manifold, this form is an even integral, and therefore
the dual conformal field theory is rational.
In this paper, we studied  a special case
of such compactifications, with the form $K_{ij}$
defining an abelian variety.
In general case,  $K_{ij}$ is an integral form, and
if $b_+ \not =b_-$, we get lattices of various signatures.
It would be interesting to understand
how these lattices can be embedded in our framework, given that  the relevant
abelian (spin) Chern-Simons theories has been recently classified   \BM.

\newsec{Conclusions and Further Directions}

In this paper we studied Hitchin-like
functionals in two dimensions.
They lead to topological theories of a special kind:
the metric is not required for constructing
the theory. Instead, it arises dynamically
from the topological data, characterized by
particular choice of the cohomologies.
We considered the cases of non-compact and compact cohomologies.
In both cases the theory generates a map  between the cohomologies
 $H^1(\Si,\IC)^{\otimes g}$ of genus $g$ Riemann surface $\Si$
 and moduli space $\CM_g$ of the complex structures on $\Si$,
in the spirit of the original Hitchin construction \refs{\Hgeom, \Hstable}.
The Hitchin parameterization of the moduli
space  in terms of the cohomologies has several useful features.
The fact that we can use simplicial complexes for
the description of cohomologies  is a natural source of the modular
group appearance.
Although explicit calculations may involve
a choice of a symplectic basis,
the action is modular invariant
and therefore provides a laboratory
for generating modular invariant objects.
Moreover, the symplectic structure on the cohomology space allows one to
perform canonical quantization of the moduli space via the Hitchin map.

\ifig\moduli{
Transport on the moduli space and the Hitchin map. }
{\epsfxsize4.0in\epsfbox{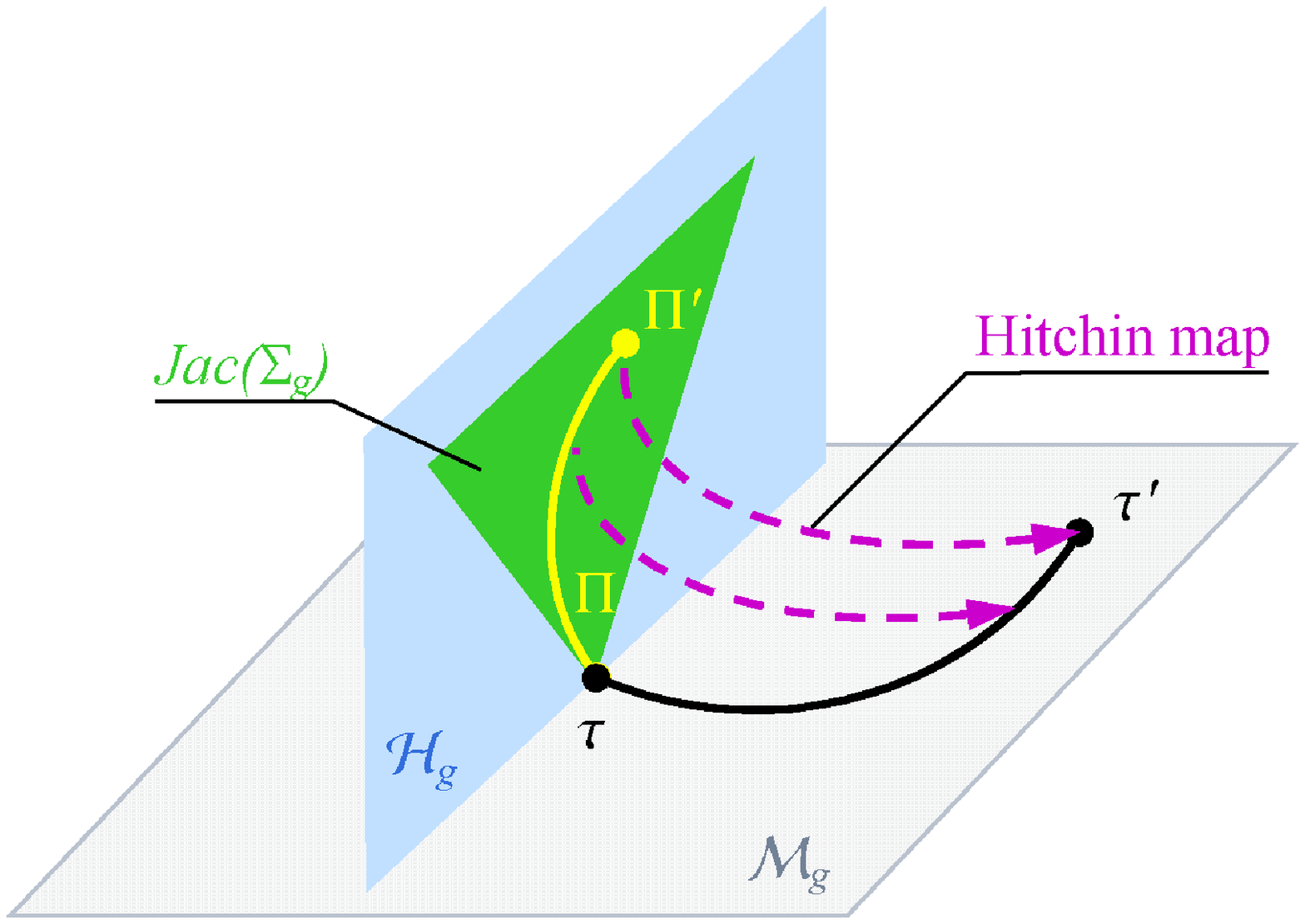}}
\noindent
The geometric picture that arises in this approach
is shown on Fig. 1. The cohomology space in question is parameterized by the
Seigel upper half-space $\CH_g$.
It can also be viewed as a space of a complex $g$-dimensional principally
polarized  abelian varieties.
The Hitchin map classically is just the Torelli map between $M_g$
and Jacobian locus $\Jac(\Si) \in \CH_g$. In the spirit of
the Kodaira-Spencer theory \BCOV, we
can start at some "background" point $\tau$ on the moduli space $\CM_g$
and study resulting quantum mechanical problem on
the Seigel upper half-space. The corresponding wave-function
 is then peaked at $\Pi =\tau$, and
classical trajectories  $\Pi \to \Pi'$ on $\CH_g$
are obtained from trajectories $\tau \to \tau'$ on $\CM_g$
via the Torelli map.

In the dual approach, we
start at some point $\Pi \in \CH_g$ and
study resulting quantum mechanical problem on
the moduli space of Riemann surfaces.
We find that in this case it is convenient to
integrate over the part of the
cohomology space given by the
complex torus $\CT(\Pi) = \IC^g/\IZ^g\oplus \Pi \IZ^g$.
Then the effective theory on $\CM_g$ can be interpreted as the abelian gauged WZW
model coupled to the two-dimensional gravity
in a special way.
Furthermore, the choice of the classical
starting points $\Pi$ is then restricted to those
that correspond to the even integral lattices.
The  classical solutions of the  gauged WZW
model correspond to the harmonic maps $\Si\to \CT$.
After coupling to the
two-dimensional gravity, variation with respect to
the complex structure
implies that these maps are holomorphic
with respect to both complex structures on $\Si(\tau) $
and $\CT(\Pi)$. This is only possible if $\CT$
is equivalent to the Jacobian of some Riemann surface, up to
the modular transformation.
In this special case
the wave function in the corresponding
quantum mechanical problem on $\CM_g$
is peaked at the attractor point \att\ $\tau = \Pi_*$.
Otherwise, the wave function is extremized at the
boundary of the moduli space $\p \CM_g$.

The probability/entropy function that we get by
squaring the wave function has  special value
at the attractor point: it is equal to the
dimension of the Hilbert space in the
 associated three-dimensional  Chern-Simons theory
with the abelian group $\CT_*$.
Therefore, the Hitchin construction
allows us to effectively organize the moduli
space of $c=2g$ RCFTs by
 introducing  canonical
index/entropy function that weights different points
on the moduli space $\CM_g$ according to
 the number of conformal blocks in corresponding RCFT.

It is widely believed that there is a vast landscape
of consistent theories of quantum gravity, that
can be realized in string theory. It was recently
suggested \Vswamp\ that
this landscape is surrounded by the huge area
of consistent looking effective theories, that cannot
be completed to a full theory, called the swampland.
On the abelian varieties side, an analog of the string landscape
is the Jacobian locus in the Siegel upper half-space,
and the "swampland" is a vast area of non-geometric points
in Siegel space,
which do not correspond to any  Riemann surface.
By extremizing the Hitchin functional,
we land on a special set of points in the Jacobian locus,
corresponding to the surfaces  admitting complex multiplication.
On the string theory side, similar phenomena occur \Moore\ if
the complex moduli of the compactification manifolds are
fixed by the attractor mechanism
\refs{\FKS, \Strominger, \CWM}.
Moreover, in both situations we have an
entropy/index weight function assigned to those
points on the moduli space.
This gives us an interesting analogy
between  the moduli space
of string compactifications and the  moduli space of
abelian varieties. Very schematically, it is shown on Fig. 2.
It would be interesting to develop this analogy further.

\ifig\land{
A similarity between the moduli spaces of
 string compactifications  and abelian varieties,
 arising if the framework of Hitchin theory. }
{\epsfxsize5.6in\epsfbox{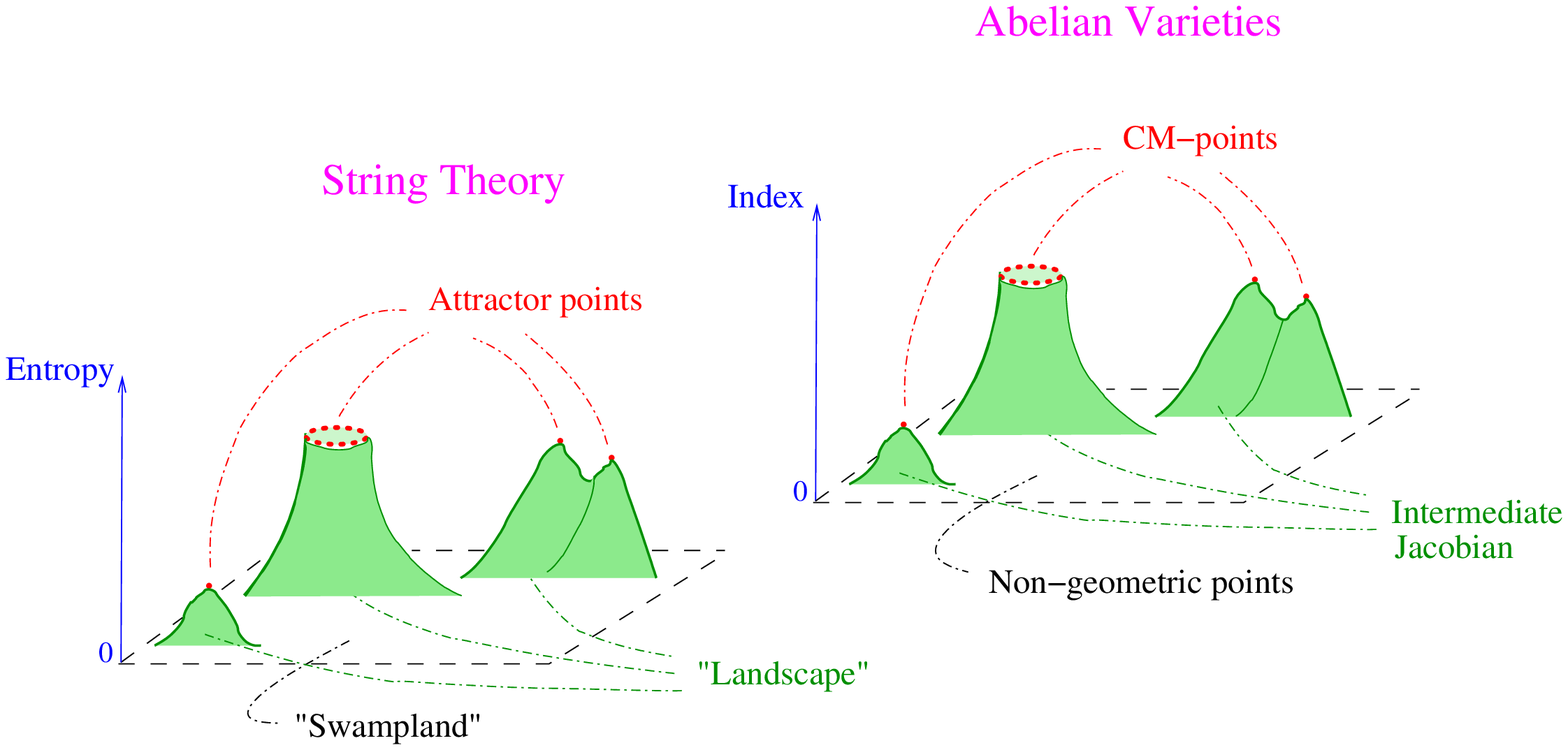}}
\noindent

In particular, it is worth mentioning that there is a direct
analog of the non-geometric  locus $\CH_g\backslash \Jac(\Si)$ in
 Siegel space  for a Calabi-Yau threefold $M$ described
in terms of the cohomologies $H^{3}(M, \IR)$.
The Hitchin theorem \Hgeom\ states that
the critical points of the functional \hcy\ on
a fixed cohomology class $[\rho]\in H^{3}(M, \IR)$ define complex structure
on a Calabi-Yau three-fold only if
there is a {\it stable} solution
$\rho_*: \ \tr K^2(\rho_*)<0$ everywhere\foot{The importance
of this condition was stressed to us
by C. Vafa.}.
From the viewpoint of the attractor equations, the boundary of the
stability region in  $H^{3}(M, \IR)$
corresponds to the black hole solutions with classically vanishing entropy.
There is no classical solutions outside of the stability region,
but it can be probed in quantum theory.
Apart form the obvious physical importance,
to  describe and classify the stability regions in $H^{3}(M, \IR)$
is a challenging mathematical question.
To the best of our knowledge, the answer to this
question is not known even in the simple case of one-parametric
Calabi-Yau threefolds. This can be thought of as the Schottky problem
for Calabi-Yau threefolds.

In this paper, we only considered
non-degenerate Riemann surfaces and concentrated on the massless degrees
of freedom. The next natural step is to
incorporate punctures and holes into the story, and
study contributions from the boundary of the moduli space.
This way, one could  control not only the
fluctuations of the geometry, but also the  change of topology.
By  concentrating on the local degrees
of freedom at the punctures it should be possible
to find a connection with the  Kodaira-Spencer theory on a
local Calabi-Yau geometries, following \ih.
Another interesting direction for further study is
incorporating supersymmetry and
considering more general curved  target spaces, for example,
non-abelian groups and $\b \g$-systems.

It is likely that the analysis of the Hitchin
functionals performed in the paper
 can be  extended  from two  to six dimensions.
The  insight that we get from studying the two-dimensional toy model \hgwzw\
is that the six-dimensional Hitchin
functional \hcy\ should be viewed as
an analog of the gauged WZW model for the seven-dimensional
Chern-Simons theory. Then
 the OSV conjecture \osv\  will have an interpretation  in terms of the
corresponding  index theorem.


\vskip 30pt

\centerline{\bf Acknowledgments}

I am grateful to C. Vafa for very useful discussions and suggestions, S. Gukov
for valuable discussions and encouragement,
and E. Witten and R. Dijkgraaf for interesting comments.
I thank A. Morozov for raising the question about  lower-dimensional
Hitchin functionals during ITEP seminar in the fall of 2001.
I also thank Irina and Vera for providing support and inspiration.

This research was supported in part by NSF grants PHY-0244821 and
DMS-0244464, {\cyr RFFI} grant 04-02-16880, and by the grant {\cyr NSh}-8004.2006.2
 for scientific schools.

\listrefs
\end